\documentclass[a4paper,11pt]{article}
\pdfoutput=1
\usepackage{jcappub}
\usepackage{epsfig}
\usepackage{amsmath}
\usepackage{amsfonts}
\usepackage{amssymb}
\usepackage{graphicx}
\usepackage{colordvi}

\title{\boldmath Astrophysical  constraints on extended gravity models}
\vspace{1cm}
\author[a,b]{Gaetano Lambiase,}
\author[c]{Mairi Sakellariadou,}
\author[a,b]{Antonio Stabile,}
\author[d]{Arturo Stabile}
\affiliation[a]{Dipartimento di Fisica ``E.R. Caianiello'', Universit\`{a} degli Studi di Salerno, via G. Paolo II, Stecca 9, I - 84084 Fisciano, Italy}
\affiliation[b]{Istituto Nazionale di Fisica Nucleare (INFN) Sezione di Napoli, Gruppo collegato di Salerno, Italy.}
\affiliation[c]{Department of Physics, King's College London,
  University of London, Strand WC2R 2LS, London, United Kingdom}
\affiliation[d]{Dipartimento di Ingegneria, Universit\`{a} del
  Sannio, Corso Garibaldi, 107 -
  82100 Benevento, Italy}

\emailAdd{lambiase@sa.infn.it}
\emailAdd{mairi.sakellariadou@kcl.ac.uk}
\emailAdd{anstabile@gmail.com}
\emailAdd{arturo.stabile@gmail.com}

\date{\today}

\abstract {
We investigate the propagation of gravitational waves in the context of fourth order gravity nonminimally coupled to a massive scalar field. Using the damping of the orbital period of coalescing stellar binary systems,  we impose constraints on the free parameters of extended gravity models. In particular, we find that
the variation of the orbital period is a function of three {\it mass} scales which depend on the free parameters of the model under consideration; we can constrain these mass scales from current observational data. 
}

\begin{document}
\begin{flushleft}
KCL-PH-TH/2015-16 \\
\end{flushleft}
\maketitle
\flushbottom

\section{Introduction}

Our Universe appears spatially flat and undergoing a period of accelerated expansion. Several observational data probe this picture~\cite{riess, ast, clo, spe,carrol,sahini}  but two unrevealed ingredients are needed in order to achieve this dynamical scenario, namely {\sl dark matter} at galactic and extragalactic scales and {\sl dark energy} at cosmological scales. The dynamical evolution of self-gravitating structures can be explained within Newtonian gravity, but a dark matter component is required in order to obtain agreement with observations~\cite{NFW}.

Lately, {\sl models of extended gravity}~\cite{report,repsergei,olmo_palatini} have been considered as a viable theoretical mechanism to explain cosmic acceleration and galactic rotation curves. In such models one extends only the geometric sector, without introducing any exotic matter. Such models may result from an effective theory of a quantum gravity formulation, which may contain additional contributions with respect to General Relativity, at galactic, extra-galactic and  cosmological scales where, otherwise, large amounts of unknown dark components are required.
In the context of {\it models of extended gravity}, 
one may consider that gravitational interaction acts differently  at different scales, whilst the robust results of General Relativity at local and solar system scales are preserved~\cite{book}.

Models of fourth order gravity have been studied in the Newtonian limit (weak-field and small velocity), as well as in the Minkowskian limit~\cite{minko}. In the former one finds  modifications of the gravitational potential, whilst in the latter one obtains massive gravitational wave modes~\cite{quadrupolo}. The weak-field limit of such proposals have to be tested against realistic self-gravitating systems. Galactic rotation curves, stellar hydrodynamics and gravitational lensing appear natural candidates as test-bed experiments~\cite{CCT, BHL, BHL1, stabile_scelza,stabstab,stabile_stabile_cap}.

Corrections to the gravitational Lagrangian were already considered by several authors~\cite{weyl_2, edd, lan,pauli,bach,buc,bic}). From a conceptual viewpoint, there is no reason \emph{a priori} to restrict the gravitational Lagrangian to a linear function of the Ricci scalar minimally coupled to matter~\cite{mag-fer-fra}.  In particular, one may consider the generalization of $f(R)$ models, where $R$ is the Ricci scalar, through generic functions containing curvature invariants such as the \emph{Ricci squared} ($R_{\alpha\beta}R^{\alpha\beta}$) or the \emph{Riemann squared} ($R_{\alpha\beta\gamma\delta}R^{\alpha\beta\gamma\delta}$), which however are not invariant due to the Gauss-Bonnet invariant $R^2-4R_{\mu\nu}R^{\mu\nu}+R_{\mu\nu\lambda\sigma}
R^{\mu\nu\lambda\sigma}$. Note that the same remark applies to the Weyl invariant $C_{\alpha\beta\gamma\delta}C^{\alpha\beta\gamma\delta}$. Hence, one may add a (massive propagating) scalar field  coupled to geometry; this leads to the {\sl scalar-tensor fourth order gravity}.

At this point, let us comment on the possibility that extended gravity theories can be plagued by pathologies, such as the appearance of ghosts (negative norm states), which could allow for negative possibilities and consequently violation of unitarity~\cite{g1,g2,g3,g4,g5,g6}.  In particular, while standard General Relativity and the Gauss-Bonnet theory have the same field content, this is not the case for the $f(R)$ gravity type and Weyl gravity. 
The former is safe, even though it does not improve the ultraviolet behaviour of the theory; $f(R)$ gravity has just an extra scalar and can be ghost free in its Newtonian limit~\cite{ghosts1}. The latter has an extra pathological spin-2 field, which however causes no problem in the low-energy regime since the effects of higher order terms give rise to small corrections to General Relativity~\cite{ghosts1,ghosts2}. In our analysis, we will consider an action motivated from noncommutative geometry within the class of a Weyl type gravity. Since, as we will discuss, this proposal will be considered in the spirit of an effective field theory, it is free from any pathologies.

Our aim here is to obtain, in the framework of post-Minkowskian approximation (weak-field limit) of  a {\sl scalar-tensor fourth order gravity}, a general solution for the wave propagation of scalar and tensor modes. The analysis will be carried out in the context of stellar binary systems. More specifically, by exploiting recent astrophysical  data on the variation of the orbital period of binary systems, we will constrain the free parameters, namely the three {\sl masses} $\{m_Y, m_R, m_\phi\}$ that characterize the scales on which higher order terms generated by the models of extended gravity become relevant. 

The outline of this paper is the following. In Sec.~\ref{STFOG}, we give the action of a {\sl scalar-tensor fourth order gravity} and write down the corresponding field equations, which we then solve in the presence of matter within the weak-field approximation. In Sec~\ref{GW_quadrupole}, we compute the gravitational wave emission from a quadrupole source, and then discuss the energy loss in Sec.~\ref{energy_loss}. Using astrophysical results on the orbital period damping,  we infer lower limits on the free parameters of {\sl scalar-tensor fourth order gravity} models studied in the literature  in Sec.~\ref{exp_cons} and Sec. \ref{STFOG_models}. We summarize our conclusions  in Sec.~\ref{conclusions}. 

\section{Scalar-Tensor Fourth Order Gravity}\label{STFOG}

Consider the action
\begin{eqnarray}\label{FOGaction}
\mathcal{S}\,=\,\int d^{4}x\sqrt{-g}\biggl[f(R,R_{\alpha\beta}R^{\alpha\beta},\phi)+\omega(\phi)\phi_{;\alpha}\phi^{;\alpha}+\mathcal{X}\mathcal{L}_{\rm m}\biggr]~,
\end{eqnarray}
where $f$ is an unspecified function of the Ricci scalar $R$, the curvature invariant $R_{\alpha\beta}R^{\alpha\beta}\,\equiv Y$ (where $R_{\mu\nu}$ is the Ricci tensor), and the scalar field $\phi$. The $\mathcal{L}_{\rm m}$ is the minimally coupled ordinary matter Lagrangian density, $\omega$ is a generic function of the scalar field $\phi$, $g$ is the determinant of the metric tensor $g_{\mu\nu}$ and $\mathcal{X}\,\equiv\,8\pi G$. We use the convention $c\,=\,1$.

In the metric approach, the field equations are obtained by varying the action (\ref{FOGaction}) with respect to $g_{\mu\nu}$, leading to
\begin{eqnarray}\label{fieldequationFOG}
f_RR_{\mu\nu}-\frac{f+\omega(\phi)\phi_{;\alpha}\phi^{;\alpha}}{2}g_{\mu\nu}-f_{R;\mu\nu}+g_{\mu\nu}\Box
f_R+2f_Y{R_\mu}^\alpha
R_{\alpha\nu}-2[f_Y{R^\alpha}_{(\mu}]_{;\nu)\alpha}
\nonumber\\+\Box[f_YR_{\mu\nu}]+[f_YR_{\alpha\beta}]^{;\alpha\beta}g_{\mu\nu}+\omega(\phi)\phi_{;\mu}\phi_{;\nu}\,=\,
\mathcal{X}\,T_{\mu\nu}~,
\end{eqnarray}
where $T_{\mu\nu}\,=\,-\frac{1}{\sqrt{-g}}\frac{\delta(\sqrt{-g}\mathcal{L}_m)}{\delta
g^{\mu\nu}}$ is the the energy-momentum tensor of matter, $f_R\,=\,\frac{\partial f}{\partial R}$, $f_Y\,=\,\frac{\partial f}{\partial Y}$ and $\Box=\nabla_\mu \nabla^\mu = {{}_{;\sigma}}^{;\sigma}$ is the D'Alembertian operator. The convention for the Ricci tensor is
$R_{\mu\nu}={R^\sigma}_{\mu\sigma\nu}$, while for the Riemann
tensor we define ${R^\alpha}_{\beta\mu\nu}=\Gamma^\alpha_{\beta\nu,\mu}+...$, with the 
affinities being the Christoffel symbols of the metric:
$\Gamma^\mu_{\alpha\beta}=\frac{1}{2}g^{\mu\sigma}(g_{\alpha\sigma,\beta}+g_{\beta\sigma,\alpha}
-g_{\alpha\beta,\sigma})$. The adopted signature is $(+---)$. The trace of the field equations
(\ref{fieldequationFOG}) reads
\begin{eqnarray}\label{tracefieldequationFOG}
f_RR+2f_YR_{\alpha\beta}R^{\alpha\beta}-2f+\Box[3
f_R+f_YR]+2[f_YR^{\alpha\beta}]_{;\alpha\beta}-\omega(\phi)\phi_{;\alpha}\phi^{;\alpha}\,=\,\mathcal{X}\,T~,
\end{eqnarray}
where $T\,=\,T^{\sigma}_{\,\,\,\,\,\sigma}$ is the trace of
energy-momentum tensor. Varying the action (\ref{FOGaction}) with respect to scalar field $\phi$ we get
\begin{eqnarray}\label{FE_SF}
2\omega(\phi)\Box\phi+\omega_\phi(\phi)\phi_{;\alpha}\phi^{;\alpha}-f_\phi\,=\,0~,
\end{eqnarray}
where $\omega_\phi(\phi)\,=\,\frac{d\omega(\phi)}{d\phi}$ and $f_\phi\,=\,\frac{\partial f}{\partial\phi}$.

We will analyze the field equations within the \emph{weak-field approximation} in a
Minkowski background $\eta_{\mu\nu}$:
\begin{equation}\label{PM_me}
g_{\mu\nu}\,\sim\,\eta_{\mu\nu}+h_{\mu\nu}
~~,~~\phi\,\sim\,\phi^{(0)}+\varphi~.
\end{equation}
We develop the function $f$ as
\begin{eqnarray}
\label{LimitFramework}
f(R,R_{\alpha\beta}R^{\alpha\beta},\phi)\,\sim &&f_{R}(0,0,\phi^{(0)})\,R+\frac{f_{RR}(0,0,\phi^{(0)})}{2}\,R^2+\frac{f_{\phi\phi}(0,0,\phi^{(0)})}{2}\,(\phi-\phi^{(0)})^2
\nonumber\\
&&+f_{R\phi}(0,0,\phi^{(0)})R\,(\phi-\phi^{(0)})+f_Y(0,0,\phi^{(0)})R_{\alpha\beta}R^{\alpha\beta}~;
\end{eqnarray}
any other possible contribution to $f$ is negligible~\cite{PRD1,PRD2,FOG_ST}. The field equations (\ref{fieldequationFOG}), (\ref{tracefieldequationFOG}) and (\ref{FE_SF}) then read
\begin{eqnarray}\label{PMfieldequationFOG2}
&&(\Box_\eta+{m_Y}^2)R_{\mu\nu}-\biggl[\frac{{m_R}^2-{m_Y}^2}{3{m_R}^2}\,
\partial^2_{\mu\nu}+\eta_{\mu\nu}\biggl(\frac{{m_Y}^2}{2}+\frac{{m_R}^2+2{m_Y}^2}{6{m_R}^2}\Box_\eta\biggr)\biggr]R
\nonumber\\
&&~~~~~~~~~~~~~~~~~~~~~~ -{m_Y}^2\,f_{R\phi}(0,0,\phi^{(0)})\,(\partial^2_{\mu\nu}-\eta_{\mu\nu}\Box_\eta)\varphi\,=\,{m_Y}^2\,\mathcal{X}\,T_{\mu\nu}~,
\nonumber\\
&&(\Box_\eta+{m_R}^2)R-3{m_R}^2\,f_{R\phi}(0,0,\phi^{(0)})\,\Box_\eta\varphi\,=\,-{m_R}^2\,\mathcal{X}\,T~,
\nonumber\\
&&(\Box_\eta+{m_\phi}^2)\varphi-f_{R\phi}(0,0,\phi^{(0)})\,R\,=\,0~,
\end{eqnarray}
where $\Box_\eta$ is the D'Alambertian operator in flat space and have made the  definitions
\begin{eqnarray}\label{mass_definition}
{m_R}^2\,&\equiv&-\frac{f_{R}(0,0,\phi^{(0)})}{3f_{RR}(0,0,\phi^{(0)})+2f_Y(0,0,\phi^{(0)})}~,\nonumber\\
{m_Y}^2\,&\equiv&\frac{f_{R}(0,0,\phi^{(0)})}{f_Y(0,0,\phi^{(0)})}~,\nonumber\\
{m_\phi}^2\,&\equiv&-\frac{f_{\phi\phi}(0,0,\phi^{(0)})}{2\omega(\phi^{(0)})}~.
\end{eqnarray}
The geometric quantities $R_{\mu\nu}$ and  $R$ are evaluated to the first order with respect to the perturbation $h_{\mu\nu}$. Note that for simplicity\footnote{We can define a new gravitational constant: $\mathcal{X}\,\rightarrow\,\mathcal{X}\,f_R(0,0,\phi^{(0)})$ and $f_{R\phi}(0,0,\phi^{0})\,\rightarrow\,f_{R\phi}(0,0,\phi^{0})\,f_R(0,0,\phi^{(0)})$.} we set $f_R(0,0,\phi^{(0)})\,=\,1$ and $\omega(\phi^{(0)})\,=\,1/2$.
The Ricci tensor in Eq.~(\ref{PMfieldequationFOG2}), in the weak-field limit, reads 
\begin{eqnarray}\label{ricci_tensor_components}
R_{\mu\nu}\,=\,h^\sigma_{\,\,\,(\mu,\nu)\sigma}-\frac{1}{2}\Box_\eta\,
h_{\mu\nu}-\frac{1}{2}h_{,\mu\nu}~,
\end{eqnarray}
where $h\,=\,{h^\sigma}_{\,\sigma}$. Using the harmonic gauge condition $g^{\rho\sigma}\Gamma^\alpha_{\,\,\,\rho\sigma}\,=\,0$ we have $h_{\mu\sigma}^{\,\,\,\,\,\,\,,\sigma}-1/2\,h_{\,,\mu}\,=\,0$, hence the Ricci tensor becomes $R_{\mu\nu}\,=\,-\frac{1}{2}\Box_\eta\,h_{\mu\nu}$. Equation~(\ref{PMfieldequationFOG2}) then reads
\begin{eqnarray}\label{PMfieldequationFOG3}
&&(\Box_\eta+{m_Y}^2)\Box_\eta h_{\mu\nu}-\biggl[\frac{{m_R}^2-{m_Y}^2}{3{m_R}^2}\,
\partial^2_{\mu\nu}+\eta_{\mu\nu}\biggl(\frac{{m_Y}^2}{2}+\frac{{m_R}^2+2{m_Y}^2}{6{m_R}^2}\Box_\eta\biggr)\biggr]
\Box_\eta h\nonumber\\
&&~~~~~~~~~~~~~ +2\,{m_Y}^2\,f_{R\phi}(0,0,\phi^{(0)})\,(\partial^2_{\mu\nu}-\eta_{\mu\nu}\Box_\eta)\varphi\,=\,-2\,{m_Y}^2\,\mathcal{X}\,T_{\mu\nu}~,
\nonumber\\
&&(\Box_\eta+{m_R}^2)\Box_\eta h+6{m_R}^2\,f_{R\phi}(0,0,\phi^{(0)})\,\Box_\eta\varphi\,=\,2{m_R}^2\,\mathcal{X}\,T~,
\nonumber\\
&&(\Box_\eta+{m_\phi}^2)\varphi+\frac{f_{R\phi}(0,0,\phi^{(0)})}{2}\,\Box_\eta h\,=\,0~.
\end{eqnarray}
The field equations (\ref{PMfieldequationFOG3}) generalize those of Ref.~\cite{GW_FOG}, where no scalar field component were considered. Note also that these equations are the weak-field limit of the model discussed in Ref.~\cite{gravito_mag}.

To solve Eq.~(\ref{PMfieldequationFOG3}) we introduce the auxiliary field $\gamma_{\mu\nu}$ such that
\begin{eqnarray}
\label{Position1}
(\Box_\eta+{m_Y}^2)\Box_\eta \gamma_{\mu\nu}&=&(\Box_\eta+{m_Y}^2)\Box_\eta h_{\mu\nu}\nonumber\\
&&-\biggl[\frac{{m_R}^2-{m_Y}^2}{3{m_R}^2}\,\partial^2_{\mu\nu}+\eta_{\mu\nu}\biggl(\frac{{m_Y}^2}{2}+\frac{{m_R}^2+2\,{m_Y}^2}{6{m_R}^2}\Box_\eta\biggr)\biggr] \Box_\eta h\nonumber
\\
&&+2\,{m_Y}^2\,f_{R\phi}(0,0,\phi^{(0)})\,(\partial^2_{\mu\nu}-\eta_{\mu\nu}\Box_\eta)\varphi~,
\end{eqnarray}
leading to
\begin{eqnarray}
\label{Position3}
h_{\mu\nu}&=& \gamma_{\mu\nu}+\biggl[\frac{{m_R}^2-{m_Y}^2}{3{m_R}^2}\,\partial^2_{\mu\nu}+\eta_{\mu\nu}\biggl(\frac{{m_Y}^2}{2}+\frac{{m_R}^2+2{m_Y}^2}{6{m_R}^2}\Box_\eta\biggr)\biggr] (\Box_\eta+{m_Y}^2)^{-1}h\nonumber
\\
&&-2\,{m_Y}^2\,f_{R\phi}(0,0,\phi^{(0)})\,(\partial^2_{\mu\nu}-\eta_{\mu\nu}\Box_\eta)(\Box_\eta+{m_Y}^2)^{-1}\Box_\eta^{-1}\varphi~.
\end{eqnarray}
Since the trace $h$ is
\begin{eqnarray}
\label{Trace3}
h= -\frac{m^2_R}{m^2_Y}(\Box_\eta+{m_Y}^2)(\Box_\eta+{m_R}^2)^{-1}\gamma -6\,{m_R}^2\,f_{R\phi}(0,0,\phi^{(0)})\,(\Box_\eta+{m_R}^2)^{-1}\varphi~,
\end{eqnarray}
Equation~(\ref{Position3}) can be written as
\begin{eqnarray}
\label{Position4}
h_{\mu\nu}&=& \gamma_{\mu\nu}-\frac{m^2_R}{m^2_Y}\biggl[\frac{{m_R}^2-{m_Y}^2}{3{m_R}^2}\,\partial^2_{\mu\nu}+\eta_{\mu\nu}\biggl(\frac{{m_Y}^2}{2}+\frac{{m_R}^2+2{m_Y}^2}{6{m_R}^2}\Box_\eta\biggr)\biggr] (\Box_\eta+{m_R}^2)^{-1}\gamma \nonumber\\
&&-2\,{m_R}^2\,f_{R\phi}(0,0,\phi^{(0)})\,\biggl[\Box_\eta^{-1}(\Box_\eta+{m_R}^2)^{-1}\partial^2_{\mu\nu}+\frac{1}{2}\eta_{\mu\nu}(\Box_\eta+{m_R}^2)^{-1}\biggl]\varphi~.
\end{eqnarray}
Using Eqs.~(\ref{Position3}), (\ref{Position4}),  Eq.~(\ref{PMfieldequationFOG3}) reads
\begin{eqnarray}\label{PMfieldequationFOG4}
(\Box_\eta+{m_Y}^2)\Box_\eta \gamma_{\mu\nu}&=&\,-2\,{m_Y}^2\,\mathcal{X}\,T_{\mu\nu}~,
\nonumber\\
(\Box_\eta+{m_Y}^2)\Box_\eta \gamma& =&\,-2\,{m_Y}^2\,\mathcal{X}\,T~,\\
(\Box_\eta+{m_\phi}^2)\varphi -3{m_R}^2f^2_{R\phi}(0,0,\phi^{(0)})\,(\Box_\eta+{m_R}^2)^{-1}\Box_\eta \varphi\,&=&\,-{m_R}^2f_{R\phi}(0,0,\phi^{(0)})\,(\Box_\eta+{m_R}^2)^{-1}\mathcal{X}\,T\,.\nonumber
\end{eqnarray}
Hence,  Eqs.~(\ref{PMfieldequationFOG3}b) and (\ref{PMfieldequationFOG3}c)
have been decoupled. Let us rewrite  Eq.~(\ref{PMfieldequationFOG4}c) as
\begin{eqnarray}\label{PMfieldequationFOG5}
(\Box_\eta+ m^2_+)(\Box_\eta+ m^2_{-})
 \varphi\,=\,-{m_R}^2f_{R\phi}(0,0,\phi^{(0)})\,\mathcal{X}\,T~,
\end{eqnarray}
where
\begin{equation}
\label{parameters}
m^2_{\pm}\equiv{m_R}^2\,w_{\pm}\bigl(\xi,\eta\bigl)~,
\end{equation}
with
\begin{eqnarray}
\label{parameters-2}w_{\pm}(\xi,\eta)&=&\frac{1-\xi+\eta^2\pm\sqrt{(1-\xi+\eta^2)^2-4\eta^2}}{2}~,
\nonumber\\
\xi&=&3{f_{R\phi}(0,0,\phi^{(0)})}^2~,\nonumber\\
\eta&=&\frac{m_\phi}{m_R}~,
\end{eqnarray}
and the constraint $\xi<1$. 

We then introduce  the auxiliary fields $\Gamma$, $\Psi$, $\Xi$ defined through
\begin{eqnarray}
\label{EquationsGammaPsiOmega}
(\Box_\eta+{m_R}^2)\Gamma\,&=&\,-{m_R}^2\,\gamma~,\nonumber\\
(\Box_\eta+{m_R}^2)\Box_\eta \Psi\,&=&\,-2\,{m_R}^2\,\varphi~,\nonumber\\
(\Box_\eta+{m_R}^2)\Xi\,&=&\,-2\,{m_R}^2\,\varphi~,
\end{eqnarray}
so that the solution (\ref{Position4}) reads
\begin{eqnarray}
\label{Position6}
h_{\mu\nu}&=&\gamma_{\mu\nu}+\frac{1}{m^2_Y}\biggl[\frac{{m_R}^2-{m_Y}^2}{3{m_R}^2}\,\partial^2_{\mu\nu}+\eta_{\mu\nu}\biggl(\frac{{m_Y}^2}{2}+\frac{{m_R}^2+2{m_Y}^2}{6{m_R}^2}\Box_\eta\biggr)\biggr] \Gamma\nonumber\\
&&+f_{R\phi}(0,0,\phi^{(0)})\,\biggl[\partial^2_{\mu\nu}\,\Psi +\frac{\Xi}{2}\eta_{\mu\nu}\biggl]~.
\end{eqnarray}
In the limit $m_R\rightarrow\infty$, $m_Y\rightarrow\infty$ and for vanishing $f_{R\phi}(0,0,\phi^{(0)})$ we recover the standard results of General Relativity (GR) since Eq.~(\ref{Position6}) reduces to $h_{\mu\nu}\,=\,\gamma_{\mu\nu}-\frac{1}{2}\eta_{\mu\nu}\gamma$.

To solve these equations we need the Green's functions (see Appendix~\ref{green_functions}); we hence consider the distributions ${\cal G}_{\rm KG,m}$, ${\cal G}_{\rm GR}$ which satisfy the equations
\begin{eqnarray}
\label{GreenFunction}
&&(\Box_\eta+ m^2){\cal G}_{\rm KG,m}(x,x^\prime)\,=\,\delta^{4}(x-x^\prime)~,
\nonumber
 \\
&&\Box_\eta {\cal G}_{\rm GR}(x,x^\prime)\,=\,\delta^{4}(x-x^\prime)~,
\end{eqnarray}
where ${\cal G}_{\rm KG,m}$ and ${\cal G}_{\rm GR}$ are the Green functions of a Klein-Gordon  field with mass $m$ (KG,m) and the standard massless models of standard General Relativity (GR), respectively. 

Due to causality, we are only interested in the retarded Green's functions, hence
\begin{eqnarray}
\label{GreenFunctionExpression}
&&{\cal G}^{\rm ret}_{\rm KG,m}(x,x^\prime)\,=\,\frac{\Theta\Bigl(t-t^\prime\Bigr)}{4\pi}\Biggl[\frac{\delta(t-t^\prime-|\textbf{x}-\textbf{x}^\prime|)}{|\textbf{x}-\textbf{x}^\prime|}-\frac{m\, {\cal J}_1(m\,\tau_{xx^\prime})}{\tau_{xx^\prime}}\Theta\Bigl(\frac{\tau_{xx^\prime}^2}{2}\Bigr)\Biggr]~, \nonumber\\
&&{\cal G}^{\rm ret}_{\rm GR}(x,x^\prime)\,=\,\frac{\Theta\Bigl(t-t^\prime\Bigr)}{4\pi}\frac{\delta(t-t^\prime-|\textbf{x}-\textbf{x}^\prime|)}{|\textbf{x}-\textbf{x}^\prime|}\,,
\end{eqnarray}
where 
\begin{equation}
 \tau_{xx^\prime}^2=(x-x^\prime)^2=(t-t^\prime)^2-|\textbf{x}-\textbf{x}^\prime|^2\,.
\end{equation}
The terms with the Dirac distribution describe the dynamics on the light cone, i.e. $t-t^\prime=|\textbf{x}-\textbf{x}^\prime|$, while the ones with the Bessel function of the first kind ${\cal J}_1(x)$ describe the dynamics interior to the light cone, i.e. $t-t^\prime>|\textbf{x}-\textbf{x}^\prime|$. We can now build the Green's functions for the auxilarly fields $\gamma_{\mu\nu}$, $\Gamma$, $\Psi$, $\Xi$ and $\varphi$ as particular combinations of the ${\cal G}^{\rm ret}_{\rm KG,m}$ and ${\cal G}^{\rm ret}_{\rm GR}$:
\begin{eqnarray}
\label{GreenFunctionRelationGammaMuNu1}
&&{\cal G}^{\rm ret}_{\gamma}(x,x^\prime)\,=\,\Theta\Bigl(t-t^\prime\Bigr)\,\Theta\Bigl(\frac{\tau_{xx^\prime}^2}{2}\Bigr)\frac{{\cal J}_1(m_Y\,\tau_{xx^\prime})}{4\pi\,m_Y\,\tau_{xx^\prime}}~,
\nonumber\\
&&{\cal G}^{\rm ret}_{\varphi}(x,x^\prime)\,=\,\frac{\Theta\Bigl(t-t^\prime\Bigr)\,\Theta\Bigl(\frac{\tau_{xx^\prime}^2}{2}\Bigr)}{4\pi\,(m^2_+-m^2_-)\,\tau_{xx^\prime}}\Biggl[  m_+\, {\cal J}_1(m_+\,\tau_{xx^\prime})-m_-\,{\cal J}_1(m_-\,\tau_{xx^\prime})\Biggl]~,
\nonumber\\
&&{\cal G}^{\rm ret}_{\Gamma}(x,x^\prime)\,=\,\frac{\Theta\Bigl(t-t^\prime\Bigr)\,\Theta\Bigl(\frac{\tau_{xx^\prime}^2}{2}\Bigr)}{4\pi\,(m^2_R-m^2_Y)\,\tau_{xx^\prime}}\Biggl[
\frac{{\cal J}_1(m_Y\,\tau_{xx^\prime})}{m_Y}-\frac{{\cal J}_1(m_R\,\tau_{xx^\prime})}{m_R}\Biggl]~,
\nonumber\\
&&{\cal G}^{\rm ret}_{\Psi}(x,x^\prime)\,=\,\frac{\Theta\Bigl(t-t^\prime\Bigr)\,\Theta\Bigl(\frac{\tau_{xx^\prime}^2}{2}\Bigr)}{4\pi\,\tau_{xx^\prime}}\Biggl[
\frac{{\cal J}_1(m_+\,\tau_{xx^\prime})}{m_+(\,m^2_--\,m^2_+)(\,m^2_R-\,m^2_+)}+\frac{{\cal J}_1(m_-\,\tau_{xx^\prime})}{m_-(\,m^2_+-\,m^2_-)(\,m^2_R-\,m^2_-)}\Bigr.\nonumber\\
\Bigl.
&&~~~~~~~~~~~~~~~+\frac{{\cal J}_1(m_R\,\tau_{xx^\prime})}{m_R(\,m^2_+-\,m^2_R)(\,m^2_--\,m^2_R)}\Biggl]~,
\nonumber\\
&&{\cal G}^{\rm ret}_{\Xi}(x,x^\prime)\,=\,-\frac{\Theta\Bigl(t-t^\prime\Bigr)\,\Theta\Bigl(\frac{\tau_{xx^\prime}^2}{2}\Bigr)}{4\pi\,\tau_{xx^\prime}}\Biggl[\frac{m_+{\cal J}_1(m_+\,\tau_{xx^\prime})}{(\,m^2_--\,m^2_+)(\,m^2_R-\,m^2_+)}+\frac{m_-{\cal J}_1(m_-\,\tau_{xx^\prime})}{(\,m^2_+-\,m^2_-)(\,m^2_R-\,m^2_-)}\Bigr.\nonumber\\
\Bigl.
&&~~~~~~~~~~~~~~~+\frac{m_R{\cal J}_1(m_R\,\tau_{xx^\prime})}{(\,m^2_+-\,m^2_R)(\,m^2_--\,m^2_R)}\Biggl]~.
\end{eqnarray}
Using Eq.~(\ref{GreenFunctionRelationGammaMuNu1}) we then derive the particular solution for the field $\gamma_{\mu\nu}$:
\begin{eqnarray}
\label{SolutionGammaMuNu}
\gamma_{\mu\nu}(t,\,\mathbf{x})\,&=&\, -\frac{m_Y\,\mathcal{X}}{2\,\pi}\,\int d^3x^\prime\,
\nonumber\\
&& ~~~~~~~~~\times\int_{-\infty}^{t-|\textbf{x}-\textbf{x}^\prime|}dt^\prime\,\frac{{\cal J}_1\bigl(m_Y\sqrt{(t-t^\prime)^2-|\textbf{x}-\textbf{x}^\prime|^2}\,\bigr)}{\sqrt{(t-t^\prime)^2-
|\textbf{x}-\textbf{x}^\prime|^2}}\,T_{\mu\nu}\bigl(t^\prime,\,\textbf{x}^\prime\bigr)~,
\end{eqnarray}
where $t-|\textbf{x}-\textbf{x}^\prime|$ is the retarded time. Introducing the variable $\tau=m_Y\sqrt{(t-t^\prime)^2-|\textbf{x}-\textbf{x}^\prime|^2}$, Eq.~(\ref{SolutionGammaMuNu}) takes the form
\begin{eqnarray}
\label{SolutionGammaMuNu2}
\gamma_{\mu\nu}(t,\,\textbf{x})&=& -\frac{m_Y\,\mathcal{X}}{2\,\pi}\,\int d^3x^\prime\,\int_{0}^{\infty}d\tau\,\frac{{\cal J}_1\bigl(\tau\bigr)}{\sqrt{\tau^2+m_Y^2|\textbf{x}-\textbf{x}^\prime|^2}}\nonumber\\
&&~~~~~~~~~~~~\times\,T_{\mu\nu}
\Biggr(t-\frac{\sqrt{\tau^2+m_Y^2|\textbf{x}-\textbf{x}^\prime|^2}}{m_Y},\,\textbf{x}^\prime\Biggr)~;
\end{eqnarray}
in the limit $m_Y\rightarrow\infty$ we obtain the standard results of General Relativity, namely
\begin{eqnarray}
\label{SolutionGammaMuNuGR}
\gamma_{\mu\nu}(t,\,\textbf{x})\,\rightarrow\,-\frac{\mathcal{X}}{2\,\pi}\,\int d^3x^\prime\,\frac{T_{\mu\nu}\bigl(t-|\textbf{x}-\textbf{x}^\prime|,\,\textbf{x}^\prime\bigr)}{|\textbf{x}-\textbf{x}^\prime|}.
\end{eqnarray}

\section{Gravitational waves emitted by a quadrupole source}\label{GW_quadrupole}

Let us assume that the sources are localized in a limited portion of  space within the neighborhood of the origin of the coordinates, namely the sources have a maximal spatial extension $|\textbf{x}^\prime_{\rm max}|$ (where $T_{\mu\nu}\neq 0$ if $|\textbf{x}|<|\textbf{x}^\prime_{\rm max}|$). If we consider the far zone limit, or radiation zone limit, i.e. $|\textbf{x}|>>\lambda>>|\textbf{x}^\prime_{\rm max}|$ where $\lambda$ is the gravitational wavelength of the waves emitted, we can consider the solution (\ref{SolutionGammaMuNu}) at a great distance from the source. In this limit (i.e. radiation zone), we set $|\textbf{x}-\textbf{x}^\prime|\approx|\textbf{x}|$ and the solution can be approximated by plane waves  having nonzero only the spatial components, i.e $\gamma_{tt}=\gamma_{ti}=0$ and $\gamma_{ij}\neq 0$. Note that in modified gravity models one has in general six different polarization states~\cite{GW_FOG}. The spatial components of $\gamma_{\mu\nu}$ can be written as
\begin{eqnarray}
\label{SolutionGammaIJFar}
\gamma_{ij}(t,\,\textbf{x})\approx -\frac{m_Y\,\mathcal{X}}{2\,\pi}\,\int_{0}^{\infty}d\tau\,\frac{{\cal J}_1\bigl(\tau\bigr)}{\sqrt{\tau^2+m_Y^2|\textbf{x}|^2}}\,\int d^3x^\prime\,T_{ij}\bigl(t-\tau_{m_Y},\,\textbf{x}^\prime\bigr)~,
\end{eqnarray}
where in general

\begin{eqnarray}
 \tau_{m}\,=\,\frac{\sqrt{\tau^2+m^2|\textbf{x}|^2}}{m}\,.
\end{eqnarray}
The spatial components of the energy-momentum tensor $T_{ij}$ are related to the quadrupole moment
\begin{eqnarray}
\label{QuadrupoleMoment}
Q_{ij}(t)\,=\,3\int d^3\mathbf{x}' \,T_{tt}(t,\,\textbf{x}')\,x'_i\,x'_j\,=\,3\int d^3\mathbf{x}' \,\rho(t,\,\textbf{x})\,x'_i\,x'_j~,
\end{eqnarray}
through the relation
\begin{eqnarray}
\label{QuadrupoleMoment2}
\int d^3\mathbf{x}'\, T_{ij}(t,\,\textbf{x}')\,=\,\frac{1}{6}\frac{d^2}{dt^2} Q_{ij}(t)\,=\,\frac{\ddot{Q}_{ij}(t)}{6}~.
\end{eqnarray}
Equation (\ref{SolutionGammaIJFar}) can be casted in the form
\begin{eqnarray}
\gamma_{ij}(\textbf{x},t)\,=\,-2\,{m_Y}\,\Upsilon^{m_Y}_{ij}(|\textbf{x}|,t)~,
\end{eqnarray}
where 
\begin{eqnarray}
\label{Upsilon}
\Upsilon^{m}_{ij}(t,\,|\textbf{x}|)\,=\,\frac{\mathcal{X}}{24\pi}\int_{0}^{\infty}d\tau\,\frac{{\cal J}_1\bigl(\tau\bigr)}{\sqrt{\tau^2+m^2|\textbf{x}|^2}}\,{\ddot Q}_{ij}\bigl(t-\tau_{m}\bigr)~.
\end{eqnarray}
Considering the trace of Eq.~(\ref{SolutionGammaMuNu}) in the radiation zone limit
\begin{eqnarray}
\label{SolutionGammaTRFar}
\gamma(t,\,\textbf{x})\approx -\frac{m_Y\,\mathcal{X}}{2\,\pi}\,\int_{0}^{\infty}d\tau\,\frac{{\cal J}_1\bigl(\tau\bigr)}{\sqrt{\tau^2+m_Y^2|\textbf{x}|^2}}\int d^3x^\prime\, T(t-\tau_{m_Y},\,\textbf{x}^\prime)~,
\end{eqnarray}
we get
\begin{eqnarray}
\label{TraceTMuNu}
\int d^3\mathbf{x}'\,T(t,\,\textbf{x}')\,=\,\eta^{\mu\nu}\int d^3\mathbf{x}'\, T_{\mu\nu}(t,\,\textbf{x}',)\,=\,M_0+\frac{\ddot{Q}(t)}{6}~,
\end{eqnarray}
where $M_0$ is the mass of the source and
$Q(t)=\eta^{ij}Q_{ij}(t)$ is the trace of the quadrupole moment (\ref{QuadrupoleMoment}). Hence, Eq.~(\ref{SolutionGammaTRFar}) becomes
\begin{eqnarray}
\label{SolutionGammaTRFar2}
\gamma(\textbf{x},t)\,&=&\,-\frac{m_Y\,M_0\,\mathcal{X}}{2\,\pi}\,\int_{0}^{\infty}d\tau\,\frac{{\cal J}_1\bigl(\tau\bigr)}{\sqrt{\tau^2+m_Y^2|\textbf{x}|^2}}
\nonumber\\
&& -\frac{m_Y\,\mathcal{X}}{12\,\pi}\,\int_{0}^{\infty}d\tau\,\frac{{\cal J}_1\bigl(\tau\bigr)}{\sqrt{\tau^2+m_Y^2|\textbf{x}|^2}}\,{\ddot Q}(t-\tau_{m_Y})~.
\end{eqnarray}
The first term on the r.h.s. of the equation above  reads
\begin{eqnarray}
\int_{0}^{\infty}d\tau\,\frac{{\cal J}_1\bigl(\tau\bigr)}{\sqrt{\tau^2+m^2|\textbf{x}|^2}}=\frac{1}{m |\textbf{x}|}\,\Bigl[ 1-e^{-m\textbf{x}} \Bigr]\,,
\end{eqnarray}
hence, it does not depend explicitly on time, so that it does not contribute to energy loss of the system. 

Using Eq.~(\ref{Upsilon}) we obtain
\begin{eqnarray}
\label{UpsilonTrace}
\Upsilon^{m}(t,\,|\textbf{x}|)\,=\eta^{ij}\Upsilon^{m}_{ij}(t,\,|\textbf{x}|)\,=\,\frac{\mathcal{X}}{24\pi}\int_{0}^{\infty}d\tau\,\frac{{\cal J}_1\bigl(\tau\bigr)}{\sqrt{\tau^2+m^2|\textbf{x}|^2}}\,{\ddot Q}(t-\tau_m)~;
\end{eqnarray}
hence the solution for the trace $\gamma$ (see Eq.~(\ref{SolutionGammaTRFar2})) takes the form
\begin{eqnarray}
\gamma(t,\,|\textbf{x}|)\,=\,-2\,{m_Y}\,\Upsilon^{m_Y}(t,\,|\textbf{x}|)~.
\end{eqnarray}
The solutions for all fields then read
\begin{eqnarray}
\begin{array}{ll}
\gamma_{ij}(t,\,|\textbf{x}|)\,=\, -2\,{m_Y}\,\Upsilon^{m_Y}_{ij}(t,\,|\textbf{x}|)~,
\\\\
\gamma(t,\,|\textbf{x}|)\,=\,-2\,{m_Y}\,\Upsilon^{m_Y}(t,\,|\textbf{x}|)~,
\\\\
\varphi(t,\,|\textbf{x}|)\,=\,-\,\frac{{m_R}^2f_{R\phi}(0,0,\phi^{(0)})}{(m^2_+-m^2_-)}\Biggl[  m_+\, \Upsilon^{m_+}(t,\,|\textbf{x}|)-m_-\Upsilon^{m_-}(t,\,|\textbf{x}|)\Biggl]~,
\\\\
\Gamma (t,\,|\textbf{x}|)\,=\,\frac{2\,{m_R}^2\,{m_Y}^2}{(m^2_R-m^2_Y)}\Biggl[\frac{\Upsilon^{m_Y}(t,\,|\textbf{x}|)}{m_Y}-\frac{\Upsilon^{m_R}(t,\,|\textbf{x}|)}{m_R}\Biggl]~,
\\\\
\Psi (t,\,|\textbf{x}|)\,=\,2\,{m_R}^4f_{R\phi}(0,0,\phi^{(0)})\,\Biggl[\frac{\Upsilon^{m_+}(t,\,|\textbf{x}|)}{m_+(\,m^2_--\,m^2_+)(\,m^2_R-\,m^2_+)}+\frac{\Upsilon^{m_-}(t,\,|\textbf{x}|)}{m_-(\,m^2_+-\,m^2_-)(\,m^2_R-\,m^2_-)}\Biggr.\\\\
~~~~~~~~~~~~~~~~~~~~~~~~~~~~~~~~~~~~~~~~~~~~\Biggl.+\frac{\Upsilon^{m_R}(t,\,|\textbf{x}|)}{m_R(\,m^2_+-\,m^2_R)(\,m^2_--\,m^2_R)}\Biggl]~,
\\\\
\Xi(t,\,|\textbf{x}|)\,=\,-2\,{m_R}^4f_{R\phi}(0,0,\phi^{(0)})\,\Biggl[\frac{m_+\Upsilon^{m_+}(t,\,|\textbf{x}|)}{(\,m^2_--\,m^2_+)(\,m^2_R-\,m^2_+)}+\frac{m_-\Upsilon^{m_-}(t,\,|\textbf{x}|)}{(\,m^2_+-\,m^2_-)(\,m^2_R-\,m^2_-)}+\frac{m_R\Upsilon^{m_R}(t,\,|\textbf{x}|)}{(\,m^2_+-\,m^2_R)(\,m^2_--\,m^2_R)}\Biggl]~.
\end{array}
\end{eqnarray}
After laborious mathematical calculations (see Appendix \ref{calculus} for details) we can rewrite the spatial components of the perturbation $h_{\mu\nu}$ given in Eq.~(\ref{Position6}), as (from Eq.~(\ref{Solution_spatial}))
\begin{eqnarray}
\label{SolutionHijFinal}
h_{ij}(|\textbf{x}|,t)&=&-2\,{m_Y}\,\Upsilon^{m_Y}_{ij}+\eta_{ij}\Biggl\{m_Y\,g_Y\,\Upsilon^{m_Y}(|\textbf{x}|,t)-m_R\,g_R\,\Upsilon^{m_R}(|\textbf{x}|,t)\nonumber\\
&&-\sum_{S=\pm}\frac{m_S\,g_S(\xi,\eta)}{3}\,\Upsilon^{m_S}(|\textbf{x}|,t)+g_R\,\Biggl[\frac{B^{m_Y}(|\textbf{x}|,t)}{m_Y}-\frac{B^{m_R}(|\textbf{x}|,t)}{m_R}\Biggr]\Biggr\}\nonumber\\
&&+\frac{2}{3}\Biggl[\frac{D^{m_Y}_{ij}(|\textbf{x}|,t)}{m_Y}+\sum_{S=\pm}g_S(\xi,\eta)\,\frac{D^{m_S}_{ij}(|\textbf{x}|,t)}{m_S}\Biggr]~.
\end{eqnarray}
This is the main result of our analysis which we will use in the following to constrain the free parameters of extended gravity models found in the literature.
\section{Energy loss}\label{energy_loss}

The rate of energy loss from a binary system, in the far-field limit, reads
\begin{eqnarray}
\label{EnergyLoss}
\frac{d{\cal E}}{dt}\,\approx\,-\frac{2\,\pi|\textbf{x}|^2}{5\,\mathcal{X}}{\dot h}_{ij}{\dot h}^{ij}~.
\end{eqnarray}
Let us make the choice
\begin{eqnarray}
\label{QuadrupoleCos}
\begin{array}{ll}
Q_{ij}\bigr(t\bigl)\,=\,{\cal Q}_{ij} \cos\bigr(\omega_{(ij)}\,t+\vartheta_{(ij)}\bigl)+{\cal Q}^0_{ij}~,
\\\\
Q\bigr(t\bigl)\,=\,{\cal Q} \cos\bigr(\nu\, t+\vartheta\bigl)+{\cal Q}^0~,
\end{array}
\end{eqnarray}
where ${\cal Q}_{ij}$ and its trace ${\cal Q}$ are the quadrupole oscillation amplitudes, ${\cal Q}_{ij}^0$ and its trace ${\cal Q}^0$ are constant terms, $\omega_{(ij)}$ and $\vartheta_{(ij)}$ are the frequencies of oscillations and the phases of the $ij$ components respectively, while $\nu$ and $\vartheta$  are the frequency and phase of the trace, respectively. All these quantities are considered to be time independent.

From Eqs.~(\ref{SolutionHijFinal}) and (\ref{EnergyLoss}),  the energy loss reads
\begin{eqnarray}
\label{EnergyLoss3}
\frac{d{\cal E}}{dt}\,&\approx\,&-\frac{m_Y^2\,\omega_{(ij)}^6{\cal Q}_{ij}{\cal Q}^{ij}\,|\textbf{x}|^2\,\mathcal{X}}{720\,\pi}F\bigl(|\textbf{x}|;m_Y;m_Y;\omega_{(ij)}\bigr)\nonumber\\
&&-\frac{\nu^6\,{\cal Q}^2\,|\textbf{x}|^2\,\mathcal{X}}{2880\,\pi}
\sum_{\{S,\,P\}}\,\zeta_{SP}\,m_S\,m_P\,F\bigl(|\textbf{x}|;m_S;m_P;\nu\bigr)~,
\end{eqnarray}
where we have averaged over time neglecting higher order terms. Note that with the notation $\sum_{\{S,\,P\}}$ we extend the sum over all possible values of $\{S,\,P\}\,=\,\{Y,\,R,\,+,\,-\}$. All quantities in Eq.~(\ref{EnergyLoss3}) are defined in Appendix~\ref{calculus}.

The model under consideration carries by itself a natural frequency scale $\omega^{\rm c}_m$ linked to the mass scales $m_R$, $m_Y$, $m_+$ and $m_-$. The $F\bigl(|\textbf{x}|;m_1;m_2;\omega\bigr)$ functions are highly oscillatory, with different behavior for $\omega>\omega^{\rm c}$ and $\omega<\omega^{\rm c}$, while for $\omega=\omega^{\rm c}$ are highly resonant~\cite{NelOchSak2010PRD}. The $\omega>\omega^{\rm c}$ case is excluded folowing a simple  heuristic argument~\cite{NelOchSak2010PRL} we highlight below. A system with $\omega>\omega^{\rm c}$ cannot decrease its orbital frequency across the lower boundary $\omega^{\rm c}$.  Since one expects all astrophysical systems to have formed from the coalescence of relatively cold, slowly moving systems, it is reasonable to suppose that at some time in the past, all binary systems had $\omega<\omega^c$. Hence, we will only analyze frequencies lower than $\omega^{\rm c}$.

For $\omega<\omega^{\rm c}$ the last function of Eq.~(\ref{FuntionF}), can be approximated by~\cite{NelOchSak2010PRD}
\begin{eqnarray}
\label{FuntionFF2}
F\bigl(|\textbf{x}|;\,m_1;\,m_2;\,\omega\bigr)\,
\approx\,
\frac{1+\Lambda\Biggl(\sqrt{m_1\,m_2}|\textbf{x}|;\,\frac{\omega}{C\,\sqrt{m_1\,m_2}}\Biggr)}{m_1\,m_2\,|\textbf{x}|^2}~,
\end{eqnarray}
where $C\approx 0.175$ is approximately constant except for $\omega\rightarrow\omega^{\rm c}$ and $\Lambda(x,\,y)\,=\,\frac{C\,{\cal J}_1(x-y)}{x(1-y)}$. From Eqs.~Eq.~(\ref{EnergyLoss3}), (\ref{FuntionFF2}) we obtain a contribution from General Relativity (GR) and one from Extended Gravity Models (EGM):
\begin{eqnarray}
\label{EnergyLoss4}
\frac{d{\cal E}}{dt}\,=\,-{\dot{\cal E}}_{\rm GR}-{\dot{\cal E}}_{\rm EGM}~,
\end{eqnarray}
where
\begin{eqnarray}
\label{EnergyLoss51}
\begin{array}{ll}
{\dot{\cal E}}_{\rm GR}\,=\,\frac{\mathcal{X}}{720	\,\pi}\Biggl(\omega_{(ij)}^6{\cal Q}_{ij}{\cal Q}^{ij}-\frac{\nu^6{\cal Q}^2}{4}\Biggr)~,
\\\\
{\dot{\cal E}}_{\rm EGM}\,=\,\frac{\mathcal{X}}{720\,\pi}\omega_{(ij)}^6{\cal Q}_{ij}{\cal Q}^{ij}\,\Lambda\Biggl(m_Y\,|\textbf{x}|;\,\frac{\omega_{(ij)}}{\omega^c_{YY}}\Biggr)+\frac{\nu^6\,{\cal Q}^2\,\mathcal{X}}{2880\,\pi}\sum_{S\,P}^{Y,\,R,\,+,\,-}\,\zeta_{SP}\,\Lambda\Biggl(\sqrt{m_S\,m_P}|\textbf{x}|;\frac{\nu}{\omega^c_{SP}}\Biggr)~,
\end{array}
\end{eqnarray}
with
\begin{eqnarray}
&&\zeta_{YY}\,=\,g_Y\,(3\,g_Y-4)\, \,,\,\,\,\zeta_{RR}\,=\,3\,g_R^2\, \,,\,\,\,\zeta_{SS}\,=\,\frac{g_S^2}{3}\, \,,\,\,\,\zeta_{YR}\,=\,(2-3\,g_Y)\,g_R~,
\nonumber\\
&&\zeta_{YS}\,=\,\frac{1}{3}\,(2-3\,g_Y)\,g_S\, \,,\,\,\,\zeta_{RS}\,=\,g_R\,g_S\, \,,\,\,\,\zeta_{\pm}\,=\,\frac{1}{3}\,g_+\,g_-~,
\end{eqnarray}
for any values in $\{S,\,P\}\,=\,\{Y,\,R,\,+,\,-\}$.

The correction to General Relativity, namely the term ${\dot {\cal E}}_{\rm EGM}$, has ten characteristic frequencies (Table~\ref{tab:tab1}). We note however that $\omega^{\rm c}_{++}=\omega^{\rm c}_{R+}$ and $\omega^{\rm c}_{--}=\omega^{\rm c}_{R-}$, and in general $\omega^{\rm c}_{+-}>\omega^{\rm c}_{++}>\omega^{\rm c}_{--}$ and $\omega^{\rm c}_{Y+}>\omega^{\rm c}_{Y-}$. Since the binary systems cannot have more frequencies than those predicted from the theory, it follows that a scalar-tensor fourth order gravity model has at most only five characteristic frequencies, i.e. $\omega^{\rm c}_{YY},\,\omega^{\rm c}_{RR},\,\omega^{\rm c}_{YR},\,\omega^{\rm c}_{--}$ and $\omega^{\rm c}_{Y-}$. Note also if the trace of the quadrupole Eq.~(\ref{QuadrupoleCos}) does not depend on time then the correction (second line of Eq.~(\ref{EnergyLoss51})) depends only on $\omega^c_{YY}$. Therefore, theories constructed without the invariant $R_{\mu\nu}R^{\mu\nu}$ will not give different values than those of General Relativity.

\begin{table}[!ht]
\centerline{
\begin{tabular}{cc}
\hline\hline
        $\omega^{\rm c}_{YY}\,=C\,m_Y$ & $\omega^{\rm c}_{R-}\,=C\,m_R\sqrt{w_-}$ \\
        $\omega^{\rm c}_{RR}\,=C\,m_R$ & \ \ \ \ $ \omega^{\rm c}_{+-}\,=C\,m_R\sqrt{w_+\,w_-}\,=C\,\sqrt{m_R\,m_\phi}$ \\
        $\omega^{\rm c}_{++}\,=C\,m_R\sqrt{w_+}$ & $\omega^c_{YR}\,=c\,\sqrt{m_Y\,m_R}$ \\
        $\omega^{\rm c}_{--}\,=C\,m_R\sqrt{w_-}$ & $\omega^{zrm c}_{Y+}\,=C\,\sqrt{m_Y\,m_R}\sqrt{w_+}$\\
        $\omega^{\rm c}_{R+}\,=C\,m_R\sqrt{w_+}$ & $ \omega^{\rm c}_{Y-}\,=C\,\sqrt{m_Y\,m_R}\sqrt{w_-}$\\
       \hline\hline
\end{tabular}}
\caption{Ten characteristic frequencies for a scalar-tensor fourth order gravity.} \label{tab:tab1}
\end{table}

The correction for the energy loss given by an EGM model takes the form
\begin{eqnarray}
\label{EnergyLossFOG2}
{\dot{\cal E}}_{\rm EGM}\,&=&\,\frac{\mathcal{X}}{720\,\pi}\Biggl[\omega_{(ij)}^6{\cal Q}_{ij}{\cal Q}^{ij}\,
+\frac{\nu^6\,{\cal Q}^2}{4}\zeta_{YY}
\Biggr]\,\Lambda\Biggl(m_Y\,|\textbf{x}|;\,\frac{\omega_{(ij)}}{\omega^{\rm c}_{YY}}\Biggr)\nonumber\\
&&~~~~~~~~+\frac{\nu^6\,{\cal Q}^2\,\mathcal{X}}{2880\,\pi}\,\Biggl[\zeta_{RR}\,\Lambda\Biggl(m_R\,|\textbf{x}|;\,\frac{\nu}{\omega^{\rm c}_{RR}}\Biggr) \nonumber
+\bigl(\zeta_{--}+\zeta_{R-}\bigr)\,\Lambda\Biggl(m_-\,|\textbf{x}|;\,\frac{\nu}{\omega^{\rm c}_{--}}\Biggr)
\nonumber\\
&&~~~~~~~~ +\zeta_{YR}\,\Lambda\Biggl(\sqrt{m_Y\,m_R}|\textbf{x}|;\,\frac{\nu}{\omega^{\rm c}_{YR}}\Biggr) +\zeta_{Y-}\,\Lambda\Biggl(\sqrt{m_Y\,m_-}|\textbf{x}|;\,\frac{\nu}{\omega^{\rm c}_{Y-}}\Biggr)
\Biggl]~.
\end{eqnarray}
As an example let us consider a pair of masses $m_1$ and $m_2$ in an elliptic binary system. For such a system, orbiting in the $(x,y)$ plane, the nonzero components of the quadrupole (\ref{QuadrupoleCos}) are
\begin{equation}
{\cal Q}_{xx}\,=\,\frac{3}{2}\,\mu\,a^2\,,\,\,\,{\cal Q}_{yy}\,=\,-\frac{3}{2}\,\mu\,b^2\,,\,\,\,{\cal Q}_{xy}\,=\,\frac{3}{2}\,\mu\,a\,b\,,\,\,\,{\cal Q}\,=\,-\frac{3}{2}\,\mu\,a^2\,e^2\,,
\end{equation}
with
\begin{equation}
\omega_{xx}\,=\,\omega_{yy}\,=\,\omega_{xy}=\,\nu\,=\,2\,\Omega~,
\end{equation}
where $\mu=\frac{m_1\,m_2}{m_1+m_2}$ is the reduced mass, $a$ and $b$ are the major and minor semiaxis, $e$ is the eccentricity and $\Omega$ is the orbital frequency. The energy loss then reads
\begin{eqnarray}
\label{EnergyLoss6}
\begin{array}{ll}
{\dot{\cal E}}_{\rm GR}\,=\,\frac{\Omega^6\,\mu^2\,\mathcal{X}}{20\,\pi}\Biggl[4\bigl(a^2+b^2\bigl)^2 -a^4\,e^2\Biggr]~,
\\\\
{\dot{\cal E}}_{\rm EGM}\,=\,\frac{\Omega^6\,\mu^2\,\mathcal{X}}{20\,\pi}\Biggr[\Bigl(4\bigl(a^2+b^2\bigl)^2 +a^4\,e^2\,\zeta_{YY}\Bigr)\,\Lambda\Biggl(m_Y\,|\textbf{x}|;\frac{2\,\Omega}{\omega^{\rm c}_{YY}}\Biggr)\\\\
~~~~~~~~~~~ +a^4\,e^2\,\Bigl[\zeta_{RR}\,\Lambda\Biggl(m_R\,|\textbf{x}|;\frac{2\,\Omega}{\omega^{\rm c}_{RR}}\Biggr)
+\bigl(\zeta_{--}+\zeta_{R-}\bigr)\,\Lambda\Biggl(m_-\,|\textbf{x}|;\,\frac{2\,\Omega}{\omega^c_{--}}\Biggr)\\\\
~~~~~~~~~~~+\zeta_{YR}\,\Lambda\Biggl(\sqrt{m_Y\,m_R}|\textbf{x}|;\,\frac{2\,\Omega}{\omega^c_{YR}}\Biggr) +\zeta_{Y-}\,\Lambda\Biggl(\sqrt{m_Y\,m_-}|\textbf{x}|;\,\frac{2\,\Omega}{\omega^{\rm c}_{Y-}}\Biggr)\Biggl]~.
\end{array}
\end{eqnarray}
These results will be used in the next section in order to constrain the three mass characterising the extended gravity model under consideration.

\section{Observational constraints}\label{exp_cons}

One has to test the observational compatibility of an extended gravity model. Hence we will study the variation of the orbital period ${\cal P}$ for binary systems due to emission of gravitational waves. The relation between the time variation of period and  the energy loss is
\begin{eqnarray}\label{1}
\frac{d{\cal P}}{dt}=\frac{\mathcal{P}^3}{4\pi I_{\rm PSR}}\frac{d{\cal E}}{dt}~,
\end{eqnarray}
where $I_{\rm PSR}$ is the pulsar's moment of inertia, normally assumed to be $10^{45}$g cm$^2$~\cite{Antoniadis}. For an elliptic binary system in the weak-field limit, one gets
\begin{eqnarray}\label{4}
\dot{{\cal P}}\approx\frac{P^3}{4\pi I_{\rm PSR}}\bigl( {\dot{\cal E}}_{\rm GR}+{\dot{\cal E}}_{\rm EGM}\bigl)=\dot{{\cal P}}_{\rm GR}+\dot{{\cal P}}_{\rm EGM}~,
\end{eqnarray}
where 
\begin{eqnarray}\begin{array}{ll}
\dot{{\cal P}}_{\rm GR}\,=\,-\frac{\Omega^6\,\mu^2\,\mathcal{P}^3\,\mathcal{X}}{80\,\pi^2\,I_{\rm PSR}}\Biggl[4\bigl(a^2+b^2\bigl)^2 -a^4\,e^2\Biggr]~,
\\\\
\dot{{\cal P}}_{\rm EGM}\,=\,-\frac{\Omega^6\,\mu^2\,P^3\,\mathcal{X}}{80\,\pi^2\,I_{\rm PSR}}\Biggr\{\biggl[4\bigl(a^2+b^2\bigl)^2 +a^4\,e^2\,\zeta_{YY}\biggr]\,\Lambda\Biggl(m_Y\,|\textbf{x}|;\,\frac{2\,\Omega}{\omega^c_{YY}}\Biggr)\\\\
~~~~~~~~+a^4\,e^2\,\Biggl[\zeta_{RR}\,\Lambda\Biggl(m_R\,|\textbf{x}|;\,\frac{2\,\Omega}{\omega^c_{RR}}\Biggr)
+\bigl(\zeta_{--}+\zeta_{R-}\bigr)\,\Lambda\Biggl(m_-\,|\textbf{x}|;\,\frac{2\,\Omega}{\omega^c_{--}}\Biggr)\\\\
~~~~~~~~ +\zeta_{YR}\,\Lambda\Biggl(\sqrt{m_Y\,m_R}|\textbf{x}|;\,\frac{2\,\Omega}{\omega^c_{YR}}\Biggr) +\zeta_{Y-}\,\Lambda\Biggl(\sqrt{m_Y\,m_-}|\textbf{x}|;\,\frac{2\,\Omega}{\omega^c_{Y-}}\Biggr)\Biggl\}~.
\end{array}
\end{eqnarray}
PSR J0348+0432 is a neutron star in a binary system with a white dwarf, with estimated massed $(2.01\pm 0.04)\,M_{\odot}$ and $(0.172\pm 0.003)\,M_{\odot}$, respectively~ \cite{Antoniadis}. This binary system has an almost circular orbit, a semi major axis $a\,=\,8.3\times 10^8\,\text{m}$ and  a short orbital period $\mathcal{P}\,=\, 2.46$ hours orbit. For these values, General Relativity leads to a significant orbital decay. In particular, the authors of Ref.~\cite{Antoniadis} obtained the constraint 

\begin{eqnarray}
 \dot{{\cal P}}_{\rm obs}/\dot{{\cal P}}_{\rm GR}\,=\,1.05\pm 0.18\,.
\end{eqnarray}
Using this result and Eq.~(\ref{4}), we get
\begin{eqnarray}\label{exp_const_circ}
\begin{array}{ll}
-0.13\leq\frac{\dot{{\cal P}}_{\rm EGM}}{\dot{{\cal P}}_{\rm GR}}\leq 0.23\,\,\,\,\,\Rightarrow\,\,\,\,\,
-0.13\leq\Lambda\Biggl(m_Y\,|\textbf{x}|;\,\frac{2\,\Omega}{\omega^c_{YY}}\Biggr) \leq 0.23~,
\\\\
\mbox{hence} \ \qquad\qquad\qquad \qquad m_Y > 5 \times 10^{-11}\,m^{-1}~.
\end{array}
\end{eqnarray}
Thus, for a binary system with a negligible circular orbit ($e\ll 1$), one can always constrain the parameter $m_Y$. To constrain the other parameters, one has to consider elliptic systems, i.e. the eccentricity must not be negligible. For example, 
for the elliptic binary system PSR B1913+16~\cite{hulse, weisberg} where the experimental eccentricity is $e\,=\,0.6$, the semi major axis $a\,=\,1.95\times 10^9\,\text{m}$, the orbital period $\mathcal{P}\,=\,7.7$ hours orbit,  and $\dot{{\cal P}}_{\rm obs}/\dot{{\cal P}}_{\rm GR}\,=\,0.997\pm 0.002$, one infers
\begin{eqnarray}\label{CostraintElliptic}
\begin{array}{ll}
-0.005\leq\frac{4\bigl(a^2+b^2\bigl)^2 +a^4\,e^2\,\zeta_{YY}}{4\bigl(a^2+b^2\bigl)^2 -a^4\,e^2}\,\Lambda\Biggl(m_Y\,|\textbf{x}|;\,\frac{2\,\Omega}{\omega^{\rm c}_{YY}}\Biggr) +\frac{a^4\,e^2}{4\bigl(a^2+b^2\bigl)^2 -a^4\,e^2}\,\Biggl[\zeta_{RR}\,\Lambda\Biggl(m_R\,|\textbf{x}|;\,\frac{2\,\Omega}{\omega^{\rm c}_{RR}}\Biggr)
\\\\
~~~~~~~~~~~~+\bigl(\zeta_{--}+\zeta_{R-}\bigr)\,\Lambda\Biggl(m_-\,|\textbf{x}|;\,\frac{2\,\Omega}{\omega^{\rm c}_{--}}\Biggr) +\zeta_{YR}\,\Lambda\Biggl(\sqrt{m_Y\,m_R}|\textbf{x}|;\,\frac{2\,\Omega}{\omega^{\rm c}_{YR}}\Biggr)\\\\
~~~~~~~~~~~~ +\zeta_{Y-}\,\Lambda\Biggl(\sqrt{m_Y\,m_-}|\textbf{x}|;\frac{2\,\Omega}{\omega^{\rm c}_{Y-}}\Biggr)\Biggr]\leq -0.001~.
\end{array}
\end{eqnarray}
We have thus obtained a relation between the characteristic masses and frequencies for a general extended gravity model. In what follows we will examine some particular extended gravity models studied in the literature. 


\section{Scalar-tensor fourth order gravity models}\label{STFOG_models}

\begin{table}[!ht]
\centering
\begin{tabular}{c|c|c}
\hline\hline\hline
Case & EGM & Mass definition \\
\hline
& & \\
A & $f(R)$ & $\begin{array}{ll}{m_R}^2\,=\,-\frac{1}{3f_{RR}(0)}\\\\{m_Y}^2\,\rightarrow\,\infty,\,\,\,{m_\phi}^2\,=\,0\\\\\xi\,=\,0,\,\,\,\eta\,=\,0\\\\m_+\,=\,m_R,\,\,\,m_-\,=\,0
\end{array}$  \\
\hline
& & \\
B & $f(R,\,R_{\alpha\beta}R^{\alpha\beta})$ & $\begin{array}{ll}{m_R}^2\,=\,-\frac{1}{3f_{RR}(0,\,0)+2f_Y(0,\,0)}\\\\{m_Y}^2\,=\,\frac{1}{f_Y(0,\,0)},\,\,\,{m_\phi}^2\,=\,0\\\\\xi\,=\,0,\,\,\,\eta\,=\,0\\\\m_+\,=\,m_R,\,\,\,m_-\,=\,0
\end{array}$  \\
\hline
&  & \\
C & $f(R,\,\phi)+\omega(\phi)\phi_{;\alpha}\phi^{;\alpha}$ & $\begin{array}{ll}{m_R}^2\,=\,-\frac{1}{3f_{RR}(0,\,\phi^{(0)})}\\\\{m_Y}^2\,\rightarrow\,\infty,\,\,\,\,{m_\phi}^2\,=\,-f_{\phi\phi}(0,\,\phi^{(0)})\\\\\xi\,=\,3{f_{R\phi}(0,\,\phi^{(0)})}^2,\,\,\,\,\eta\,=\,\frac{m_\phi}{m_R}\\\\m_\pm\,=\,\sqrt{\frac{1-\xi+\eta^2\pm\sqrt{(1-\xi+\eta^2)^2-4\eta^2}}{2}}\,m_R
\end{array}$  \\
\hline
&  & \\
D & $c_0R+c_1\,R\,\phi+f(\phi)+\omega(\phi)\phi_{;\alpha}\phi^{;\alpha}$ & $\begin{array}{ll}{m_R}^2\rightarrow\,\infty\,\,,\,{m_Y}^2\rightarrow\,\infty,\,\,\,\,{m_\phi}^2\,=\,-f_{\phi\phi}(\phi^{(0)})\\\\\xi\,=\,3\,{c_1}^2,\,\,\,\,\eta\,\rightarrow\,0\\\\m_+\,\rightarrow\,\infty,\,\,\,m_-\,\rightarrow\,\frac{m_\phi}{\sqrt{1-3\,{c_1}^2}}
\end{array}$  \\
\hline
&  & \\
E & $f(R,\,R_{\alpha\beta}R^{\alpha\beta},\phi)+\omega(\phi)\phi_{;\alpha}\phi^{;\alpha}$ & $\begin{array}{ll}{m_R}^2\,=\,-\frac{1}{3f_{RR}(0,\,0,\,\phi^{(0)})+2f_Y(0,\,0,\,\phi^{(0)})}\\\\{m_Y}^2\,=\,\frac{1}{f_Y(0,0,\phi^{(0)})},\,\,\,\,{m_\phi}^2\,=\,-f_{\phi\phi}(0,\,0,\,\phi^{(0)})\\\\\xi\,=\,3{f_{R\phi}(0,0,\phi^{(0)})}^2,\,\,\,\,\eta\,=\,\frac{m_\phi}{m_R}\\\\m_\pm\,=\,\sqrt{\frac{1-\xi+\eta^2\pm\sqrt{(1-\xi+\eta^2)^2-4\eta^2}}{2}}\,m_R
\end{array}$  \\
\hline\hline\hline
\end{tabular}
\caption{\label{tab:tab2}Here $f_R(0,\,0,\,\phi^{(0)})\,=\,1$ and $\omega(\phi^{(0)})\,=\,1/2$ and for the case D we set also $c_0+c_1\phi^{(0)}\,=\,1$.}
\end{table}

Let us consider case A of Table~\ref{tab:tab2}; the only interesting quantity (see Eq.~(\ref{mass_definition})) is $m_R$. For the system PSR B1913+16,
Eq.~(\ref{CostraintElliptic}) implies
\begin{eqnarray}
\label{CostraintEllipticF(R)}
-0.005\leq \frac{a^4\,e^2\,\Lambda\Biggl(m_R\,|\textbf{x}|;\,\frac{2\,\Omega}{\omega^c_{RR}}\Biggr)}{4\bigl(a^2+b^2\bigl)^2 -a^4\,e^2}\leq -0.001\,\Rightarrow m_R\,\gtrsim\, 3\times 10^{-9}\,\text{m}^{-1}~.
\end{eqnarray}
In general, one can consider the polynomial expression

\begin{eqnarray}
f(R)\,=\,R+\alpha\,R^2+\sum_{n\,=\,3}^{N}\alpha_n\,R^n\,.
\end{eqnarray} 
Note however that the characteristic scale $m_R$ is only generated by the $R^2$-term. An interesting model of $f(R)$-theories is that of Starobinsky $f(R)\,=\,R-R^2/R_0$~\cite{staro}, for which $m_R^2=R_0/6$, hence using Eq.~(\ref{CostraintEllipticF(R)}) we get $R_0\gtrsim\, 5.4\times 10^{-19}\,\text{m}^{-2}$.

To generalize the previously result we must include the  curvature invariant $R_{\mu\nu}R^{\mu\nu}$. For case B of Table~\ref{tab:tab2} we consider the general class of $f(R,\,R_{\alpha\beta}R^{\alpha\beta})$-theories and their characteristic scales $m_R$ and $m_Y$. Using Eqs.~(\ref{exp_const_circ}) and (\ref{CostraintElliptic}) we obtain
 \begin{eqnarray}
 m_Y\,\gtrsim\, 5\times 10^{-11}\,\text{m}^{-1}\ \ , \ \ 
m_R\,\gtrsim\, 1.15\times 10^{-9}\,\text{m}^{-1}~.
 \end{eqnarray}
This class of theories includes the case of a Weyl square type model, \emph{i.e.} $C_{\mu\nu\rho\sigma}C^{\mu\nu\rho\sigma}\,=\,2R_{\mu\nu}R^{\mu\nu}-\frac{2}{3}R^2$, where there is only one characteristic scale $m_R\,\rightarrow\,\infty$.

The same argumentation is also valid for the scalar-tensor case of theory, for which in the Newtonian limit (see case D in Table \ref{tab:tab2}) the more general expression (\ref{LimitFramework}) becomes

\begin{eqnarray}
\biggl(1-\phi^{(0)}\sqrt{\frac{\xi}{3}}\biggr)R+\sqrt{\frac{\xi}{3}}\,R\,\phi-\frac{{m_\phi}^2}{2}\,(\phi-\phi^{(0)})^2~.
\end{eqnarray}
Thus, for the most general Scalar-Tensor (ST) theory in the Newtonian limit, one can consider the model\footnote{With the condition $\alpha_0+\alpha_1\,\phi^{(0)}\,=\,1$.}

\begin{eqnarray}\label{scalar-tensor}
f_{\rm ST}(R,\phi)\,=\,c_0\,R+c_1\,R\,\phi-\frac{1}{2}{m_\phi}^2\,(\phi-\phi^{(0)})^2+\frac{1}{2}\phi_{,\alpha}\phi^{,\alpha}~.
\end{eqnarray}
Since for this case $m_R\rightarrow \infty $, $m_Y\rightarrow \infty$, $\xi\,=\,3\,{c_1}^2$, $\eta\,\rightarrow\,0$, $m_+\rightarrow \infty$ and  $m_-\,=\,\frac{m_\phi}{\sqrt{1-3{c_1}^2}}$, we obtain from Eq.~(\ref{CostraintElliptic}) 

\begin{eqnarray}
\label{CostraintEllipticTensorScalar}
-0.005\leq\frac{a^4\,e^2}{4\bigl(a^2+b^2\bigl)^2 -a^4\,e^2}\,\frac{2\,{c_1}^2(3-6\,{c_1}^2)}{(1-3\,{c_1}^2)^2}\,\Lambda\Biggl(\frac{m_\varphi\,|\textbf{x}|}{\sqrt{1-3\,{c_1}^2}};\,\frac{2\,\Omega}{\omega^{\rm c}_{--}}\Biggr)\leq -0.001~.
\end{eqnarray}

As a special case of  a scalar-tensor fourth order gravity model (case E) we consider NonCommutative Spectral Geometry (NCSG)~\cite{connes_1,connes_2}, for which at a cutoff scale (set as the Grand Unification scale) the purely gravitational part of the action coupled to the Higgs ${\bf H}$ reads~\cite{ccm}
\begin{equation}\label{eq:action1} 
{\cal S}_{\rm NCSG}\,=\,\int d^4x\sqrt{-g}\biggl[\frac{R}{2{\kappa_0}^2}+ \alpha_0\,C_{\mu\nu\rho\sigma}C^{\mu\nu\rho\sigma} 
+\tau_0 R^\star R^\star+\frac{{\bf H}_{;\alpha}{\bf H}^{;\alpha}}{2}-{\mu_0}^2\,
{\bf H}^2-\frac{R\,{\bf H}^2}{12} +\lambda_0\,{\bf H}^4\biggr]\,,
\end{equation} 
where $R^\star R^\star$ is the topological term that integrates to the Euler characteristic, hence nondynamical.
Since the square of the Weyl tensor can be expressed in terms of $R^2$ and $R_{\mu\nu}R^{\mu\nu}$ as $C_{\mu\nu\rho\sigma}C^{\mu\nu\rho\sigma}\,=\,2R_{\mu\nu}R^{\mu\nu}-\frac{2}{3}R^2$, the NCSG action is a particular case of action (\ref{FOGaction}).

At this point, let us briefly discuss the behavior of the gravitational part of the action Eq.~(\ref{eq:action1}) above. We emphasise that Eq.~(\ref{eq:action1})  is the result of a perturbative expansion of the bosonic spectral action ${\rm Tr}(\chi(D^2/\Lambda^2))$, with $\chi$ a cutoff action, $D$ the Dirac operator and $\Lambda$ a constant scale up to which the theory is valid; usually taken as the Grand Unified Theories scale~\cite{ccm}. The higher derivative terms that are quadratic in curvature can be written in the form~\cite{ccm,donogue}
\begin{equation}\label{eq:action2} 
\,\int d^4x\sqrt{-g}\biggl[\frac{1}{2\eta}\,C_{\mu\nu\rho\sigma}C^{\mu\nu\rho\sigma} 
-\frac{\omega}{3\eta}R^2+\frac{\theta}{\eta}E\biggr]\,,
\end{equation} 
with $E=R^\star R^\star$ the topological term which is the integrand in the Euler characteristic. Renormalisation Group Equations (RGE) determine the running of the coefficients $\eta, \omega, \theta$ of the higher derivative terms in the action Eq.~(\ref{eq:action2}). The (weak) low energy constraints on the coefficients of the quadratic curvature terms $R_{\mu\nu}R^{\mu\nu}$ and $R^2$ impose that these coefficients should not exceed $10^{74}$~\cite{donogue}, which is indeed the case for the action Eq.~(\ref{eq:action2}) as RGE analysis has shown~\cite{ccm}. 

Using Eqs.~(\ref{exp_const_circ}) and (\ref{CostraintElliptic}), we can constrain the parameter $\alpha_0$, which corresponds to a restriction on the particle physics at unification. 
We thus obtain
\begin{equation}
\alpha_0 \leq 10^{20} {\rm m}^{2}~,
\end{equation}
which is rather weak but can in principle be improved once further data of nearby pulsars are available.

The parameter $\alpha_0$ has been constrained in the past using either pulsar measurements~\cite{NelOchSak2010PRL}, Gravity Probe B or torsion balance experiments~\cite{LamSakSta}. Here we have extended the original analysis of Ref.~\cite{NelOchSak2010PRL} for the case of pulsars with an elliptical orbit. Let us note that the strongest constraint on $\alpha_0$, namely {$\alpha_0<10^{-8}\,{\rm m}^2$,  was obtained~\cite{LamSakSta} using torsion balance measurements.

\section{Conclusions}\label{conclusions}

In the context of {\sl extended gravity models} and in particular within the class of {\sl scalar-tensor fourth order gravity}, we have studied the energy loss of stellar binary systems. 
In general, the models we have considered depend on four parameters, as
\begin{eqnarray}
\label{LimitFramework_02}
f(R,R_{\alpha\beta}R^{\alpha\beta},\phi)+\omega(\phi)\phi_{;\alpha}\phi^{;\alpha}=
&&\biggl(1-\phi^{(0)}\,\sqrt{\frac{\xi}{3}}\biggr)R-\frac{2{m_R}^2+{m_Y}^2}{6{m_R}^2{m_Y}^2}\,R^2+\frac{R_{\alpha\beta}R^{\alpha\beta}}{{m_Y}^2}
\nonumber\\\\\nonumber
&&+\sqrt{\frac{\xi}{3}}\,R\,\phi+\frac{\phi_{;\alpha}\phi^{;\alpha}}{2}-\frac{{m_\phi}^2}{2}(\phi-\phi^{(0)})^2~,
\end{eqnarray}
where we have set $f_R(0,\,0,\,\phi^{(0)})\,=\,1$ and $\omega(\phi^{(0)})\,=\,1/2$. Note that $m_R^{-1}$, $m_Y^{-1}$, $m_\phi^{-1}$ determine the characteristic lenghts of propagation, $\xi$ is the coupling constant between the background geometry and the scalar field, while the background value $\phi^{(0)}$ of the scalar field does not determine directly the dynamics.

To constrain the free parameters we considered the nearly circular binary system PSR J0348+0432 and  found  $m_Y > 5 \times 10^{-11}\,m^{-1}$. Considering the elliptic binary system PSR B1913+16 with eccentricity $e\,=\,0.6$ we have constrained all three free parameters. Choosing a particular scalar-tensor fourth order gravity  model (\ref{LimitFramework_02}) we obtained a lower value for $m_R$ and $m_\phi$, namely $m_R\,\gtrsim\, 3\times 10^{-9}\,\text{m}^{-1}$ and $m_\phi\,\gtrsim\, 1.3\times 10^{-11}\,\text{m}^{-1}$, respectively. One may be able to set stronger constraints by considering systems which are closer.
It is worth noting that for circular binary systems there are no corrections in the case of a pure $f(R)$ gravity with respect to General Relativity.

\appendix

\section{Green functions for a ScalarTensor Fourth Order Gravity}\label{green_functions}

The complete set of equations replacing the field equations (\ref{PMfieldequationFOG3}) for $h_{\mu\nu}$, $\varphi$ and Eq.~(\ref{EquationsGammaPsiOmega}) for the auxiliarly fields  $\gamma_{\mu\nu}$, $\Gamma$, $\Psi$, $\Xi$ are
\begin{eqnarray}
\label{EquationsGammaPhi}
\begin{array}{ll}
(\Box_\eta+{m_Y}^2)\Box_\eta\gamma_{\mu\nu}\,=\,-2\,{m_Y}^2\,\mathcal{X}\,T_{\mu\nu}~,\\\\
(\Box_\eta+{m_Y}^2)\Box_\eta\gamma\,=\,-2\,{m_Y}^2\,\mathcal{X}\,T~,\\\\
(\Box_\eta+ {m_+}^2)(\Box_\eta+ {m_{-}}^2)\varphi\,=\,-{m_R}^2f_{R\phi}(0,0,\phi^{(0)})\,\mathcal{X}\,T~,\\\\
(\Box_\eta+{m_R}^2)(\Box_\eta+{m_Y}^2)\Box_\eta\Gamma\,=\,2\,{m_R}^2\,{m_Y}^2\,\mathcal{X}\,T~,\\\\
(\Box_\eta+ {m_+}^2)(\Box_\eta+ {m_-}^2)(\Box_\eta+{m_R}^2)\Box_\eta\Psi\,=\,2\,{m_R}^4f_{R\phi}(0,0,\phi^{(0)})\,\mathcal{X}\,T~,\\\\
(\Box_\eta+ {m_+}^2)(\Box_\eta+ {m_-}^2)(\Box_\eta+{m_R}^2)\Xi\,=\,2\,{m_R}^4f_{R\phi}(0,0,\phi^{(0)})\,\mathcal{X}\,T~,
\end{array}
\end{eqnarray}
where we have four characteristic lengths (${m_R}^{-1}$, ${m_Y}^{-1}$, ${m_+}^{-1}$, ${m_-}^{-1}$), which we assume all different and real. The first two (${m_R}^{-1}$, ${m_Y}^{-1}$) are generated by the geometry, while the last two (${m_+}^{-1}$, ${m_-}^{-1}$)  are lengths resulting from the interaction between geometry and the scalar field $\varphi$.

The solutions of Eq.~(\ref{EquationsGammaPhi}) can be expressed in terms of Green functions as 
\begin{eqnarray}
\begin{array}{ll}
\gamma_{\mu\nu}(x)\,=\,-2\,{m_Y}^2\,\mathcal{X}\,\int d^4x^\prime{\cal G}_{\gamma}(x,x^\prime)\,T_{\mu\nu}(x^\prime)~,
\\\\
\gamma(x)\,=\,-2\,{m_Y}^2\,\mathcal{X}\,\int d^4x^\prime{\cal G}_{\gamma}(x,x^\prime)\,T(x^\prime)~,
\\\\
\varphi(x)\,=\,-\,{m_R}^2f_{R\phi}(0,0,\phi^{(0)})\,\mathcal{X}\,\int d^4x^\prime{\cal G}_{\varphi}(x,x^\prime)\,T(x^\prime)~,
\\\\
\Gamma (x)\,=\,\,2\,{m_R}^2\,{m_Y}^2\,\mathcal{X}\,\int d^4x^\prime{\cal G}_{\Gamma}(x,x^\prime)\,T(x^\prime)~,
\\\\
\Psi (x)\,=\,\,2\,{m_R}^4f_{R\phi}(0,0,\phi^{(0)})\,\mathcal{X}\,\int d^4x^\prime{\cal G}_{\Psi}(x,x^\prime)\,T(x^\prime)~,
\\\\
\Xi (x)\,=\,\,2\,{m_R}^4f_{R\phi}(0,0,\phi^{(0)})\,\mathcal{X}\,\int d^4x^\prime{\cal G}_{\Xi}(x,x^\prime)\,T(x^\prime)~,
\end{array}
\end{eqnarray}
with the Green functions fixed by 
\begin{eqnarray}
\begin{array}{ll}
(\Box_\eta+{m_Y}^2)\Box_\eta{\cal G}_{\gamma}(x,x^\prime)\,=\,\delta^{4}(x-x^\prime)~,
\\\\
(\Box_\eta+ m^2_+)(\Box_\eta+ m^2_{-}){\cal G}_{\varphi}(x,x^\prime)\,=\,\delta^{4}(x-x^\prime)~,
\\\\
(\Box_\eta+{m_R}^2)(\Box_\eta+{m_Y}^2)\Box_\eta{\cal G}_{\Gamma}(x,x^\prime)\,=\,\delta^{4}(x-x^\prime)~,
\\\\
(\Box_\eta+ m^2_+)(\Box_\eta+ m^2_-)(\Box_\eta+{m_R}^2)\Box_\eta{\cal G}_{\Psi}(x,x^\prime)\,=\,\delta^{4}(x-x^\prime)~,
\\\\
(\Box_\eta+ m^2_+)(\Box_\eta+ m^2_-)(\Box_\eta+{m_R}^2){\cal G}_{\Xi}(x,x^\prime)\,=\,\delta^{4}(x-x^\prime)~,
\end{array}
\end{eqnarray}
where $\delta^{4}(x-x^\prime)$ is a four-dimensional Dirac distribution in flat space-time. To find the analytical dependence of the Green functions it can be shown that in Fourier space they are linear combination of only ${\cal G}_{\rm KG,m}$ and ${\cal G}_{\rm GR}$, which satisfy the second order equations
\begin{eqnarray}
\begin{array}{ll}
(\Box_\eta+ m^2){\cal G}_{\rm KG,m}(x,x^\prime)\,=\,\delta^{4}(x-x^\prime)~,
 \\\\
\Box_\eta {\cal G}_{\rm GR}(x,x^\prime)\,=\,\delta^{4}(x-x^\prime)~,
\end{array}
\end{eqnarray}
with solutions 
\begin{eqnarray}
\begin{array}{ll}
{\cal G}^{\rm ret}_{\rm KG,m}(x,x^\prime)\,=\,\frac{\Theta\Bigl(t-t^\prime\Bigr)}{4\pi}\Biggl[\frac{\delta(t-t^\prime-|\textbf{x}-\textbf{x}^\prime|)}{|\textbf{x}-\textbf{x}^\prime|}-\frac{m\, {\cal J}_1(m\,\tau_{xx^\prime})}{\tau_{xx^\prime}}\Theta\Bigl(\frac{\tau_{xx^\prime}^2}{2}\Bigr)\Biggr]~, \\
\nonumber
{\cal G}^{\rm ret}_{\rm GR}(x,x^\prime)\,=\,\frac{\Theta\Bigl(t-t^\prime\Bigr)}{4\pi}\frac{\delta(t-t^\prime-|\textbf{x}-\textbf{x}^\prime|)}{|\textbf{x}-\textbf{x}^\prime|}~,
\end{array}
\end{eqnarray}
where $\tau_{xx^\prime}^2=(x-x^\prime)^2=(t-t^\prime)^2-|\textbf{x}-\textbf{x}^\prime|^2$. 

Hence the Green's functions ${\cal G}^{\rm ret}_{\gamma}, {\cal G}^{\rm ret}_{\varphi}, {\cal G}^{\rm ret}_{\Psi}$ and ${\cal G}^{\rm ret}_{\Xi}$ are 
\begin{eqnarray}
\begin{array}{ll}
{\cal G}^{\rm ret}_{\gamma}(x,x^\prime)= \frac{1}{m^2_Y}\Biggl[ {\cal G}^{\rm ret}_{\rm GR}(x,x^\prime)-{\cal G}^{\rm ret}_{{\rm KG},m_Y}(x,x^\prime) \Biggl]=\Theta\Bigl(t-t^\prime\Bigr)\,\Theta\Bigl(\frac{\tau_{xx^\prime}^2}{2}\Bigr)\frac{{\cal J}_1(m_Y\,\tau_{xx^\prime})}{4\pi\,m_Y\,\tau_{xx^\prime}}~,
\\\\
{\cal G}^{\rm ret}_{\varphi}(x,x^\prime)=\frac{1}{m^2_+-m^2_-}\Biggl[ {\cal G}^{\rm ret}_{{\rm KG},m^2_-}(x,x^\prime)-{\cal G}^{\rm ret}_{KG,m^2_+}(x,x^\prime) \Biggl]\\\\
~~~~~~~~~~~~~~~~~~=\frac{\Theta\Bigl(t-t^\prime\Bigr)\,\Theta\Bigl(\frac{\tau_{xx^\prime}^2}{2}\Bigr)}{4\pi\,(m^2_+-m^2_-)\,\tau_{xx^\prime}}\Biggl[  m_+\, {\cal J}_1(m_+\,\tau_{xx^\prime})-m_-\,{\cal J}_1(m_-\,\tau_{xx^\prime})\Biggl]~,
\\\\
{\cal G}^{\rm ret}_{\Gamma}(x,x^\prime)= \frac{1}{m^2_R-m^2_Y}\Biggl[\frac{m^2_R-m^2_Y}{m^2_R\,m^2_Y}{\cal G}^{\rm ret}_{\rm GR}(x,x^\prime)+\frac{1}{m^2_R}{\cal G}^{\rm ret}_{{\rm KG},m_R}(x,x^\prime)-\frac{1}{m^2_Y}{\cal G}^{\rm ret}_{{\rm KG},m_Y}(x,x^\prime)\Biggl]
\\\\
\qquad\qquad\,\,\,\,=\frac{\Theta\Bigl(t-t^\prime\Bigr)\,\Theta\Bigl(\frac{\tau_{xx^\prime}^2}{2}\Bigr)}{4\pi\,(m^2_R-m^2_Y)\,\tau_{xx^\prime}}\Biggl[
\frac{{\cal J}_1(m_Y\,\tau_{xx^\prime})}{m_Y}-\frac{{\cal J}_1(m_R\,\tau_{xx^\prime})}{m_R}\Biggl]~,
\\\\
{\cal G}^{\rm ret}_{\Psi}(x,x^\prime)=\frac{{\cal G}^{\rm ret}_{\rm GR}(x,x^\prime)}{m^2_+\,m^2_-\,m^2_R}-\frac{{\cal G}^{\rm ret}_{KG,m_+}(x,x^\prime)}{m^2_+(\,m^2_--\,m^2_+)(\,m^2_R-\,m^2_+)}-\frac{{\cal G}^{\rm ret}_{KG,m_-}(x,x^\prime)}{m^2_-(\,m^2_+-\,m^2_-)(\,m^2_R-\,m^2_-)} -\frac{{\cal G}^{\rm ret}_{{\rm KG},m_R}(x,x^\prime)}{m^2_R(\,m^2_+-\,m^2_R)(\,m^2_--\,m^2_R)}
\\\\
\qquad\qquad\,\,\,\,=\frac{\Theta\Bigl(t-t^\prime\Bigr)\,\Theta\Bigl(\frac{\tau_{xx^\prime}^2}{2}\Bigr)}{4\pi\,\tau_{xx^\prime}}\Biggl[\frac{{\cal J}_1(m_+\,\tau_{xx^\prime})}{m_+(\,m^2_--\,m^2_+)(\,m^2_R-\,m^2_+)}+\frac{{\cal J}_1(m_-\,\tau_{xx^\prime})}{m_-(\,m^2_+-\,m^2_-)(\,m^2_R-\,m^2_-)}
+\frac{{\cal J}_1(m_R\,\tau_{xx^\prime})}{m_R(\,m^2_+-\,m^2_R)(\,m^2_--\,m^2_R)}\Biggl]~,
\\\\
{\cal G}^{\rm ret}_{\Xi}(x,x^\prime)= \frac{{\cal G}^{\rm ret}_{KG,m_+}(x,x^\prime)}{(\,m^2_--\,m^2_+)(\,m^2_R-\,m^2_+)}+\frac{{\cal G}^{\rm ret}_{{\rm KG},m_-}(x,x^\prime)}{(\,m^2_+-\,m^2_-)(\,m^2_R-\,m^2_-)} +\frac{{\cal G}^{\rm ret}_{{\rm KG},m_R}(x,x^\prime)}{(\,m^2_+-\,m^2_R)(\,m^2_--\,m^2_R)}
\\\\
\qquad\qquad\,\,\,\,=-\frac{\Theta\Bigl(t-t^\prime\Bigr)\,\Theta\Bigl(\frac{\tau_{xx^\prime}^2}{2}\Bigr)}{4\pi\,\tau_{xx^\prime}}\Biggl[\frac{m_+{\cal J}_1(m_+\,\tau_{xx^\prime})}{(\,m^2_--\,m^2_+)(\,m^2_R-\,m^2_+)}+\frac{m_-{\cal J}_1(m_-\,\tau_{xx^\prime})}{(\,m^2_+-\,m^2_-)(\,m^2_R-\,m^2_-)}
+\frac{m_R{\cal J}_1(m_R\,\tau_{xx^\prime})}{(\,m^2_+-\,m^2_R)(\,m^2_--\,m^2_R)}\Biggl]~.
\end{array}
\end{eqnarray}
\section{Mathematical aspects of the metric components $h_{ij}$}\label{calculus}

The spatial components of the perturbation $h_{\mu\nu}$ (\ref{Position6}) can be expressed as
\begin{eqnarray}\label{metric_compo_ij}
h_{ij}(t,\,|\textbf{x}|)=&&-2\,{m_Y}\,\Upsilon^{m_Y}_{ij}(t,\,|\textbf{x}|)+\frac{1}{3m_Y}\Bigl[2\,\partial^2_{ij}+\eta_{ij}\,\text{H}\,\Bigr]\Upsilon^{m_Y}(t,\,|\textbf{x}|)-\eta_{ij}\,\Pi\,\Upsilon^{m_R}(t,\,|\textbf{x}|)\nonumber
\\\\
&&+\sum_{S=\pm}\frac{g_S(\xi,\eta)}{3\,m_S}\Bigl[2\partial^2_{ij}-\eta_{ij}\,m_S^2\Bigl]\Upsilon^{m_S}(t,\,|\textbf{x}|)~,\nonumber
\end{eqnarray}
where
\begin{eqnarray}
\begin{array}{ll}
\text{H}\,=\,\frac{\Bigl[9\, m_R^2m_Y^2+3\,(2\,m_R^2+m_Y^2)\Box\Bigr]}{(m_R^2-m_Y^2)}~,
\\\\
\Pi\,=\,\frac{\Bigl[(2\,m_R^2+m_Y^2)(\Box +m_R^2)\Bigr]}{m_R\,(m_R^2-m_Y^2)}~,
\\\\
g_\pm(\xi,\eta)\,=\,\frac{\xi}{\bigl[w_\mp(\xi,\eta)-w_\pm(\xi,\eta)\bigr]\bigl[1-w_\pm(\xi,\eta)\bigr]}~,
\end{array}
\end{eqnarray}
and $\Upsilon^m_{ij}(t,\,\mathbf{x})$  and $\Upsilon^m(t,\,\mathbf{x})$ are defined in Eqs.~(\ref{Upsilon}), (\ref{UpsilonTrace}). The derivatives of $\Upsilon^m(t,\,\mathbf{x})$ are
\begin{eqnarray}
\label{UpsilonTraceDerivative}
\partial_{\mu}\Upsilon^{m}(t,\,|\textbf{x}|)\,&=&\,\frac{\mathcal{X}}{24\pi}\int_{0}^{\infty}d\tau\,{\cal J}_1\bigl(\tau\bigr)\,\Biggl[-\frac{m^2\,\partial_{\mu}(|\textbf{x}|^2){\ddot Q}(t-\tau_m)}{2\bigr[\tau^2+m^2|\textbf{x}|^2\bigr]^{3/2}}+\frac{\partial_{\mu}{\ddot Q}(t-\tau_m)}{\sqrt{\tau^2+m^2|\textbf{x}|^2}}\Biggr]~,
\nonumber\\
\partial^2_{\mu\nu}\Upsilon^{m}(t,\,|\textbf{x}|)\,&=&\,\frac{\mathcal{X}}{24\pi}\int_{0}^{\infty}d\tau\,{\cal J}_1\bigl(\tau\bigr)\,
\Biggl\{\frac{3\,m^4}{4}\frac{\partial_{\mu}(|\textbf{x}|^2)\partial_{\nu}(|\textbf{x}|^2){\ddot Q}(t-\tau_m)}{\bigr[\tau^2+m^2|\textbf{x}|^2\bigr]^{5/2}}-\frac{m^2\,\partial^2_{\mu\nu}(|\textbf{x}|^2){\ddot Q}(t-\tau_m)}{2\,\bigr[\tau^2+m^2|\textbf{x}|^2\bigr]^{3/2}}
\nonumber\\
&&-\frac{m^2}{2}\biggl[ \frac{\partial_{\mu}(|\textbf{x}|^2)\,\partial_{\nu}{\ddot Q}(t-\tau_m)+\partial_{\nu}(|\textbf{x}|^2)\,\partial_{\mu}{\ddot Q}(t-\tau_m)}{\bigr[\tau^2+m^2|\textbf{x}|^2\bigr]^{3/2}}\biggr]+\frac{\partial^2_{\mu\nu}{\ddot Q}(t-\tau_m)}{\sqrt{\tau^2+m^2|\textbf{x}|^2}}\Biggr\}~,
\nonumber\\
\partial^2_{ij}\Upsilon^{m}(t,\,|\textbf{x}|)\,&=&\,\frac{\mathcal{X}}{24\pi}\int_{0}^{\infty}d\tau\,{\cal J}_1\bigl(\tau\bigr)\,\Biggl\{\frac{m^2\,{\ddot Q}(t-\tau_m)}{\bigr[\tau^2+m^2|\textbf{x}|^2\bigr]^{3/2}}\biggl[ \frac{3\,m^2\,x_i\,x_j}{\bigr[\tau^2+m^2|\textbf{x}|^2\bigr]}-\delta_{ij}\biggr]
\nonumber\\
&&+\frac{m\,{\ddot Q}^{\prime}(t-\tau_m)}{\bigr[\tau^2+m^2|\textbf{x}|^2\bigr]}\biggl[ \frac{2\,m^2\,x_i\,x_j}{\bigr[\tau^2+m^2|\textbf{x}|^2\bigr]}-\delta_{ij}\biggr]+\frac{m^2\,x_i\,x_j\,{\ddot Q}^{\prime\prime}(t-\tau_m)}{\bigr[\tau^2+m^2|\textbf{x}|^2\bigr]^{3/2}}\Biggr\}~,
\nonumber\\
\Box\Upsilon^{m}(t,\,|\textbf{x}|)\,&=&\,\frac{\mathcal{X}}{24\pi}\int_{0}^{\infty}d\tau\,{\cal J}_1\bigl(\tau\bigr)\,\Biggl[\frac{3\,m^2\,{\ddot Q}(t-\tau_m)\,\tau^2}{\bigr[\tau^2+m^2|\textbf{x}|^2\bigr]^{5/2}}
+\frac{3\,m\,{\ddot Q}^{\prime}(t-\tau_m)\,\tau^2}{\bigr[\tau^2+m^2|\textbf{x}|^2\bigr]^2}
\nonumber\\
&&-\frac{m^2\,|\textbf{x}|^2\,{\ddot Q}^{\prime\prime}(t-\tau_m)}{\bigr[\tau^2+m^2|\textbf{x}|^2\bigr]^{3/2}}
+\frac{{\ddddot Q}(t-\tau_m)}{\sqrt{\tau^2+m^2|\textbf{x}|^2}}\Biggr]~.
\end{eqnarray}
Equations (\ref{UpsilonTraceDerivative}c) and  (\ref{UpsilonTraceDerivative}d) can be approximated by considering only the terms scaling as $1/|\mathbf{x}|$; the other terms scale as $1/|\mathbf{x}|^n$ with $n\,>\,1$. Thus, we have
\begin{eqnarray}
\label{UpsilonTraceDerivative3}
\begin{array}{ll}
\partial^2_{ij}\Upsilon^{m}(t,\,|\textbf{x}|)\,\approx\,\frac{\mathcal{X}}{24\pi}\int_{0}^{\infty}d\tau\,{\cal J}_1\bigl(\tau\bigr)\,\frac{\partial^2_{ij}{\ddot Q}(t-\tau_m)}{\sqrt{\tau^2+m^2|\textbf{x}|^2}}\\\\
~~~~~~~~~~~~~~~~~\approx\frac{m^2\,x_i\,x_j\,\mathcal{X}}{24\pi}\int_{0}^{\infty}d\tau\,{\cal J}_1\bigl(\tau\bigr)\,\frac{{\ddot Q}^{\prime\prime}(t-\tau_m)}{\bigr[\tau^2+m^2|\textbf{x}|^2\bigr]^{3/2}}\,\equiv D^{m}_{ij}(|\textbf{x}|,t)~,
\\\\
\Box\Upsilon^{m}(t,\,|\textbf{x}|)\,\approx\,\frac{\mathcal{X}}{24\pi}\int_{0}^{\infty}d\tau\,{\cal J}_1\bigl(\tau\bigr)\frac{{\ddddot Q}(t-\tau_m)}{\sqrt{\tau^2+m^2|\textbf{x}|^2}}\equiv B^{m}(|\textbf{x}|,t)~,
\end{array}
\end{eqnarray}
and Eq.~(\ref{metric_compo_ij}) reads
\begin{eqnarray}
\label{Solution_spatial}
\begin{array}{ll}
h_{ij}(t,\,|\textbf{x}|)\,=\,-2\,{m_Y}\,\Upsilon^{m_Y}_{ij}+\eta_{ij}\Biggl\{m_Y\,g_Y\,\Upsilon^{m_Y}(t,\,|\textbf{x}|)-m_R\,g_R\,\Upsilon^{m_R}(t,\,|\textbf{x}|)
\\\\ \qquad\qquad\qquad\,\,\,
-\sum_{S=\pm}\frac{m_S\,g_S(\xi,\eta)}{3}\,\Upsilon^{m_S}(t,\,|\textbf{x}|)
+g_R\,\Biggl[\frac{B^{m_Y}(|\textbf{x}|,t)}{m_Y}-\frac{B^{m_R}(|\textbf{x}|,t)}{m_R}\Biggr]\Biggr\}
\\\\ \qquad\qquad\qquad\,\,\,+\frac{2}{3}\Biggl[\frac{D^{m_Y}_{ij}(t,\,|\textbf{x}|)}{m_Y}+\sum_{S=\pm}g_S(\xi,\eta)\,\frac{D^{m_S}_{ij}(t,\,|\textbf{x}|)}{m_S}\Biggr]~,
\end{array}
\end{eqnarray}
where
\begin{eqnarray}
\label{gFunction}
g_Y\,=\,\frac{3\,m_R^2}{m_R^2-m_Y^2}\ \ \ \ ,\ \ \ \ g_R\,=\,\frac{2\,m_R^2+m_Y^2}{m_R^2-m_Y^2}~.
\end{eqnarray}
The time derivatives of $\Upsilon^m_{ij}(t,\,|\mathbf{x}|)$, $B^{m}(t,\,|\textbf{x}|)$, $D^{m}_{ij}(t,\,|\textbf{x}|)$ needed to calculate the energy loss in Eq.~(\ref{EnergyLoss}) are 
\begin{eqnarray}
\label{UpsilonIJTimeDerivative}
\begin{array}{ll}
{\dot\Upsilon^{m}}_{ij}(t,\,|\textbf{x}|) \,=\, \frac{\omega_{(ij)}^3\,{\cal Q}_{ij}\,\mathcal{X}}{24\,\pi}\,\Biggl[ \sin\bigr(\omega_{(ij)}\,t+\vartheta_{(ij)}\bigl)f^c_1\biggl(m|\textbf{x}|;\,\frac{\omega_{(ij)}}{m}\biggr)- \cos\bigr(\omega_{(ij)}\,t+\vartheta_{(ij)}\bigl)f^s_1\biggl(m|\textbf{x}|;\,\frac{\omega_{(ij)}}{m}\biggr) \Biggr]~,
\\\\
{\dot\Upsilon^{m}}(t,\,|\textbf{x}|)\,=\,\frac{\nu^3\,{\cal Q}\,\mathcal{X}}{24\,\pi}\,\Biggl[\sin\bigr(\nu\,t+\vartheta\bigl)f^c_1\biggl(m|\textbf{x}|;\,\frac{\nu}{m}\biggr)- \cos\bigr(\nu\,t+\vartheta\bigl)f^s_1\biggl(m|\textbf{x}|;\,\frac{\nu}{m}\biggr) \Biggr]~,
\\\\
{\dot B^{m}}(t,\,|\textbf{x}|)\,=\,-\frac{\nu^5\,{\cal Q}\,\mathcal{X}}{24\,\pi}\,\Biggl[\sin\bigr(\nu\,t+\vartheta\bigl)f^c_1\biggl(m|\textbf{x}|;\,\frac{\nu}{m}\biggr)- \cos\bigr(\nu\,t+\vartheta\bigl)f^s_1\biggl(m|\textbf{x}|;\,\frac{\nu}{m}\biggr) \Biggr]~,
\\\\
{\dot D^{m}}_{ij}(t,\,|\textbf{x}|)\,=\,-\frac{\nu^5\,m^2\,x_i\,x_j\,{\cal Q}\,\mathcal{X}}{24\,\pi}\,\Biggl[\sin\bigr(\nu\,t+\vartheta\bigl)f^c_3\biggl(m|\textbf{x}|;\,\frac{\nu}{m}\biggr)- \cos\bigr(\nu\,t+\vartheta\bigl)f^s_3\biggl(m|\textbf{x}|;\,\frac{\nu}{m}\biggr) \Biggr]~,
\end{array}
\end{eqnarray}
with the definitions
\begin{eqnarray}
\label{FuntionF}
\begin{array}{ll}
f^c_n\bigl(x;z\bigr)\,\equiv\,\int_{0}^{\infty}d\tau\,\frac{{\cal J}_1\bigl(\tau\bigr)\,\cos\bigr(z\,\sqrt{\tau^2+x^2}\bigl)}{\bigr(\sqrt{\tau^2+x^2}\,\bigl)^n}~,
\\\\
f^s_n\bigl(x;z\bigr)\,\equiv\,\int_{0}^{\infty}d\tau\,\frac{{\cal J}_1\bigl(\tau\bigr)\,\sin\bigr(z\,\sqrt{\tau^2+x^2}\bigl)}{\bigr(\sqrt{\tau^2+x^2}\,\bigl)^n}~,
\\\\
F\bigl(|\textbf{x}|;m_1;m_2;\omega\bigr)\,\equiv\,f^c_1\biggl(m_1|\textbf{x}|;\frac{\omega}{m_1}\biggr)f^c_1\biggl(m_2|\textbf{x}|;\frac{\omega}{m_2}\biggr)+f^s_1\biggl(m_1|\textbf{x}|;\frac{\omega}{m_1}\biggr)f^s_1\biggl(m_2|\textbf{x}|;\frac{\omega}{m_2}\biggr)~.
\end{array}
\end{eqnarray}

\end{document}